\RequirePackage{fix-cm}
\documentclass[smallextended]{svjour3}       % onecolumn (second format)
\smartqed  % flush right qed marks, e.g. at end of proof
\usepackage{graphicx}
\usepackage[margin=1.5in]{geometry}
\usepackage[export]{adjustbox}
\usepackage{url}
\usepackage{color}
\begin{document}

\title{Photo- and Electrocouplings of Nucleon Resonances}

\author{Victor I. Mokeev \and Daniel S. Carman \\ (for the CLAS Collaboration)}

\institute{Victor I. Mokeev \at 
           Jefferson Laboratory, 12000 Jefferson Ave., Newport News, VA 23602, USA
           \email{mokeev@jlab.org}
           \and
           Daniel S. Carman \at
           Jefferson Laboratory, 12000 Jefferson Ave., Newport News, VA 23602, USA
           \email{carman@jlab.org}}

\date{Received: date / Accepted: date}

\maketitle

\begin{abstract}
Advances in the exploration of the spectrum and structure of the excited states of the nucleon from experiments with electromagnetic probes 
on proton targets are presented. Impressive progress has been achieved in the studies of exclusive meson photoproduction in experiments with 
continuous electron beams and with detectors of almost $4\pi$ acceptance. The high-quality data, coupled with the advances in the amplitude 
analyses of exclusive photo- and hadroproduction data, allow for the observation of several long-awaited new baryon states known previously 
as the ``missing" resonances. Studies of exclusive meson electroproduction in the resonance region with the CLAS detector at JLab have 
provided the dominant part of the available world information on exclusive meson electroproduction observables. These data offer unique 
information on the structure of most well-established excited nucleon states in the mass region up to 1.8~GeV in terms of the
evolution of their electroexcitation amplitudes with momentum transfer of the virtual photon. We discuss the impact of these results on the 
insight into the strong interaction dynamics that underlie the generation of the full spectrum of nucleon resonances of distinctively 
different structure. These results shed light on the emergence of hadron mass, which is one of the most important and still open problems in 
the Standard Model. The extension of the nucleon resonance studies in the experiments of the 12-GeV era at JLab with the CLAS12 detector are
outlined.
\end{abstract}

\section{Introduction}
\label{intro}

Studies of the spectrum and structure of the excited states of the nucleon (generically referred to here as $N^*$s) represent an important 
part of the effort to explore the strong interaction dynamics of quantum chromodynamics (QCD) in the regime of large (comparable with unity) 
running coupling. This regime of what is termed strong QCD underlies the generation of hadrons (mesons and baryons) as bound systems 
of quarks and gluons~\cite{Brodsky:2020vco,Burkert:2019bhp}. The experimental results on the $N^*$ spectrum shed light on the approximate 
symmetries of strong QCD that are relevant for the generation of these states. The full $N^*$ spectrum, including those states already 
observed and those that are still to be discovered, defines the rate for the transition from the primordial deconfined mixture of quarks 
and gluons into the hadron gas phase that took place within the first microseconds after the Big Bang~\cite{baz2014a,bur2020}. In this phase
transition, quark-gluon confinement emerged and the masses of hadrons were generated in close connection with the dynamical breaking of the
approximate chiral symmetry of QCD. These features define the essence of the strong QCD regime that makes the studies of the $N^*$ spectrum 
a compelling experimental program to explore the emergence of hadron matter in the early Universe and the evolution towards its contemporary 
status. 

Studies of $N^*$ structure from the data on the nucleon resonance electroexcitation amplitudes, the $\gamma_vpN^*$ electrocouplings, 
as a function of the $Q^2$, the squared four-momentum transfer of the virtual photon to the target proton, offer a unique opportunity to 
explore the many facets of strong QCD involved in the generation of $N^*$s of different quantum numbers and structural features
\cite{Burkert:2019bhp,fbs-carman,azbu12,barab20212}. This information has provided critically needed input for the understanding of strong 
QCD dynamics and its emergence from the QCD Lagrangian. Any theoretical approach for the description of hadron structure with a connection 
to QCD is required to describe the structure of both the ground and excited states of the nucleon within a common framework. Studies of the
$\gamma_vpN^*$ electrocouplings address key open problems of the Standard Model on the emergence of hadron mass (EHM) and on the nature of
quark-gluon confinement in connection with dynamical chiral symmetry breaking~\cite{cdrob20,cui20}.

In this lecture the advances of the past decade in the exploration of the spectrum and structure of the $N^*$ states are reviewed. The impact 
of these studies on understanding strong QCD dynamics is discussed.

\section{Manifestation of Nucleon Resonances in Inclusive Hadro-, Photo-, and Electroproduction Data}
\label{inclusive}

The global analyses of the data on inclusive and semi-inclusive electron scattering off nucleons, Drell-Yan lepton-pair production, 
$W$/$Z$-boson asymmetries, and $\gamma$+jet cross sections revealed the structure of the ground state nucleon as a bound system of a
large number of current quarks and gluons (partons) in relativistic motion~\cite{moff2021,accard2016}. These constituents are always 
confined inside the nucleon interior and have never been observed as free particles. Such a complex composite system should possess a rich 
spectrum of excited states. 

%%%%%%%%%%%%%%%%%%%%%%%%%%%%%%%%%%%%%%%%%%%%%%%%%%%%%%%%%%%%%%%%%%%%%%%%%%%%%%%%%%%%%%%%%%%%%%%%%%%%%%%%%%%%%%%%%%%%%%%%%%%%%%%%%%%%%%%%%%%%%%
\begin{figure}
  \raisebox{11mm}{\includegraphics[width=0.53\textwidth,height=5.0cm]{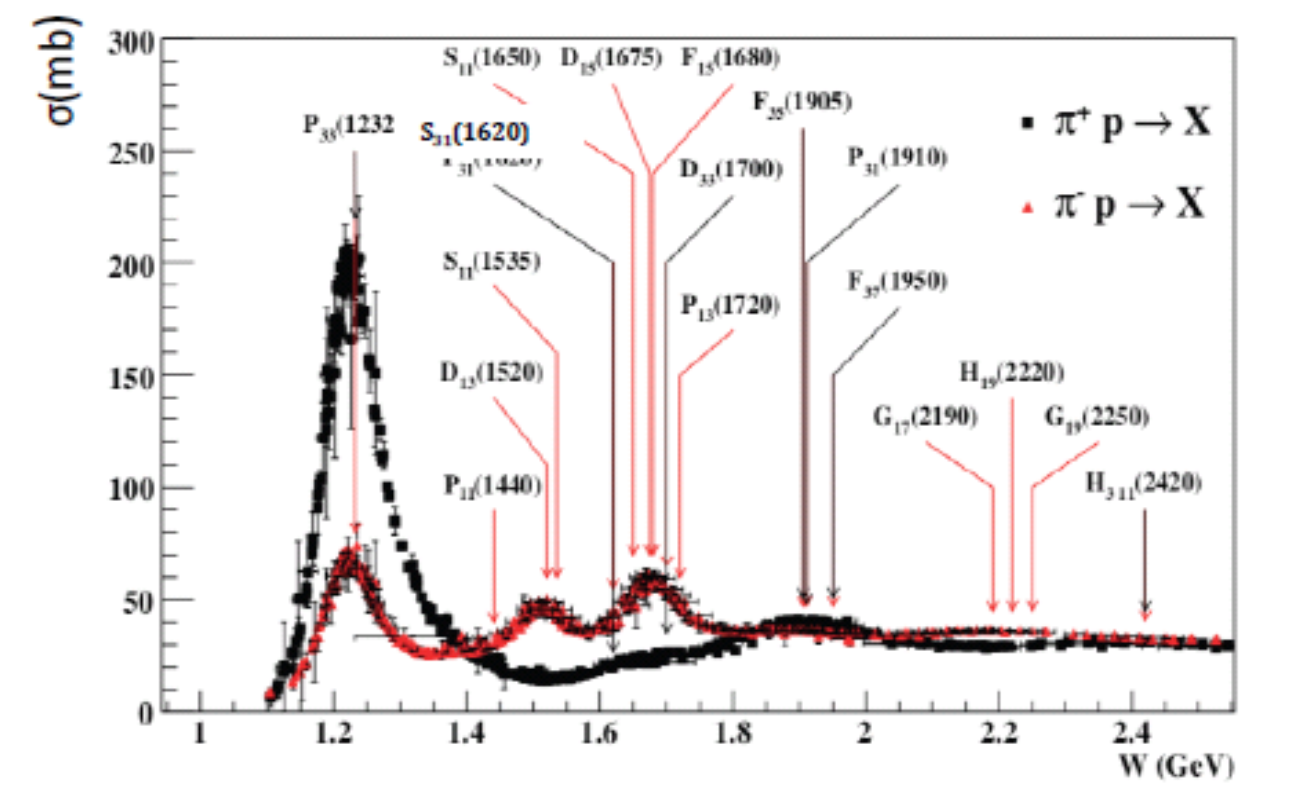}}
  \raisebox{0mm}{\includegraphics[width=0.45\textwidth]{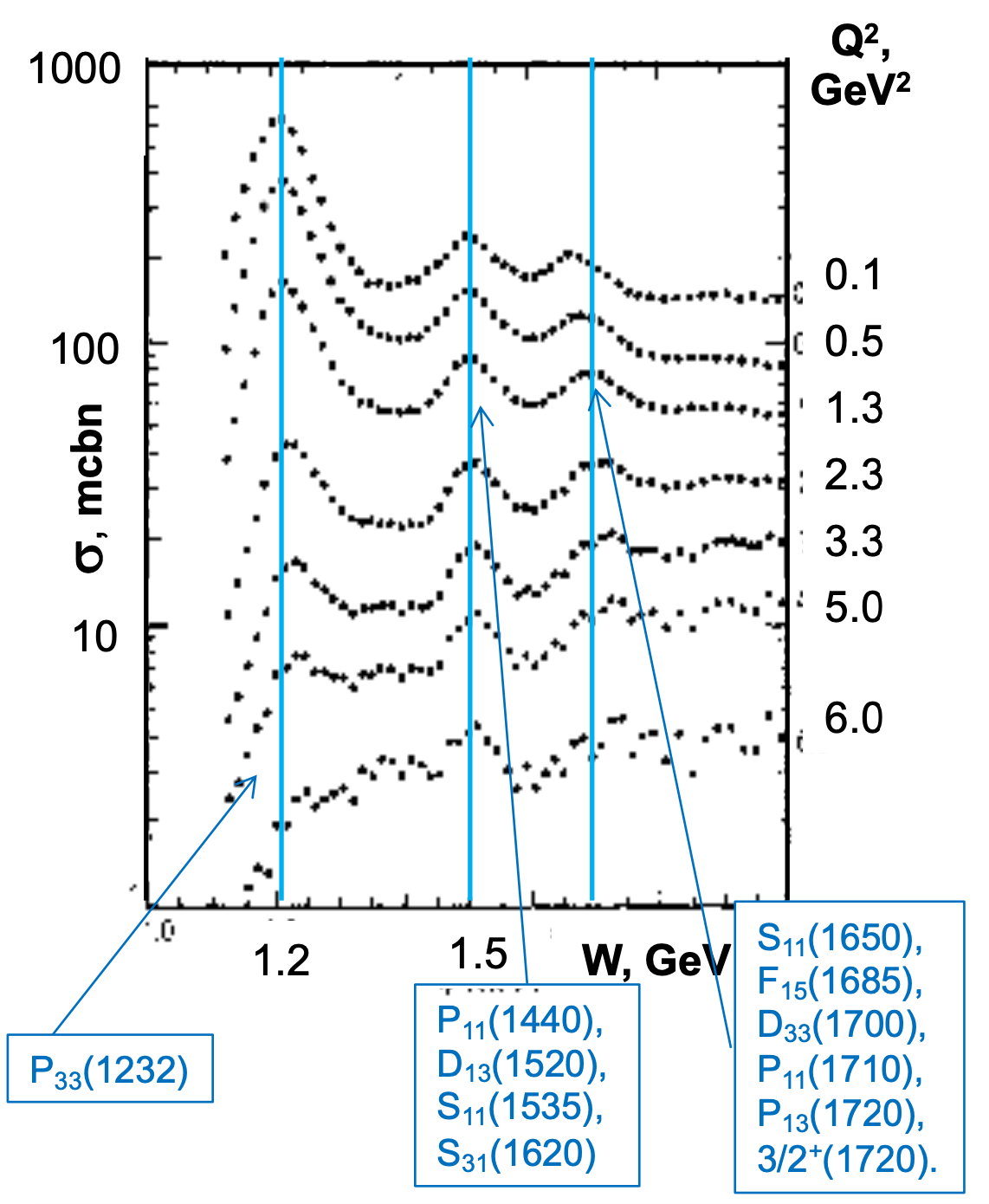}}
\caption{(Left) Inclusive $\pi N$ cross section as a function of the invariant mass $W$ of the final state hadrons fully integrated over the 
final state kinematic variables~\cite{crede2013}. (Right) Fully integrated $\gamma_v p \to X$ inclusive cross sections as a function of $W$ 
and $Q^2$~\cite{foster1983}. The nucleon resonances contributing into the first, second, and the third resonance regions are shown in both 
figures.}
\label{incl_hadelectr}
\end{figure}
%%%%%%%%%%%%%%%%%%%%%%%%%%%%%%%%%%%%%%%%%%%%%%%%%%%%%%%%%%%%%%%%%%%%%%%%%%%%%%%%%%%%%%%%%%%%%%%%%%%%%%%%%%%%%%%%%%%%%%%%%%%%%%%%%%%%%%%%%%%%%%

\begin{figure}[htbp]
  \includegraphics[width=0.5\textwidth,height=5.0cm]{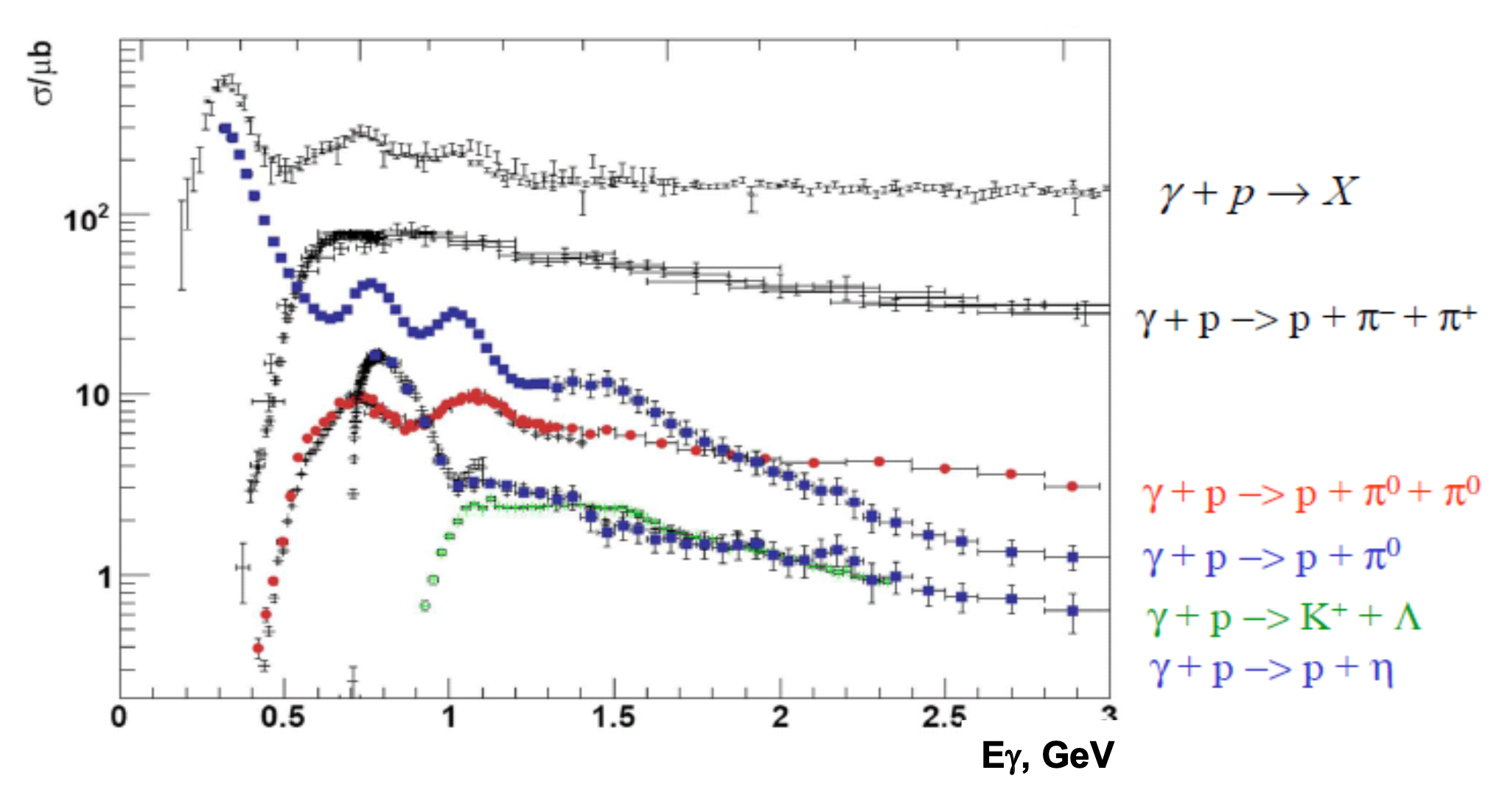}
  \includegraphics[width=0.48\textwidth,height=5.0cm]{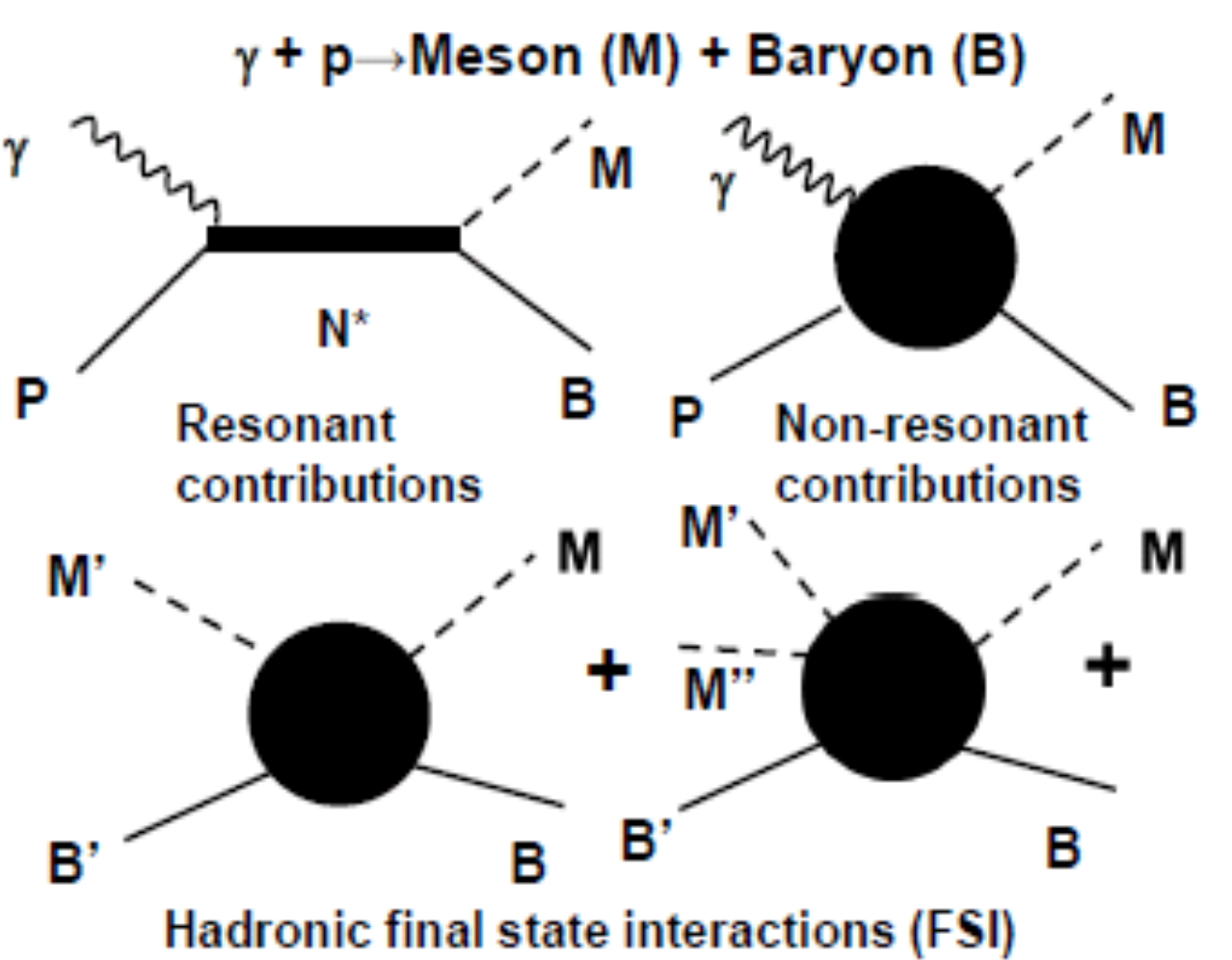}
\caption{(Left) Fully integrated exclusive meson photoproduction cross sections in the resonance region as a function of the incident photon 
energy~\cite{beck2017}. (Right) Schematic representation of the meson-baryon photoproduction amplitudes: the sum of the resonant and 
non-resonant contributions is shown in the top row and the hadronic initial/final state interactions with the open meson-baryon $M'B'$ and 
$M'M''B$ channels are shown in the bottom row.}
\label{integ_ex_photo}
\end{figure}

The contributions from nucleon resonances have been observed in total pion-nucleon cross sections and in inclusive photo- and 
electroproduction off proton targets~\cite{pdg,foster1983}. In Figs.~\ref{incl_hadelectr} and \ref{integ_ex_photo} these cross sections are 
shown as a function of the invariant mass $W$ of the final hadron system or the incident photon energy $E_\gamma$ for photoproduction. The 
inclusive $(e,e'X)$ electroproduction cross sections were measured over a broad range of $Q^2$. In the $W$-dependence of the measured cross
sections, resonance-like peaks are clearly seen. Their maxima in both the hadro- and electroproduction data are located at nearly the same
$W$-values: 1.23~GeV (the first resonance region), 1.52~GeV (the second resonance region), and 1.70~GeV (the third resonance region). These 
features suggest contributions from $s$-channel resonances in $\pi N$ and $\gamma_vp$ collisions. The $\pi N$ channels are 
sensitive to those resonances with substantial decays to the $\pi N$ final states. Instead, the photo- and electroproduction channels are 
sensitive to the contributions from those $N^*$ states with sizable decays not only to $\pi N$ but also to other final hadron states such 
as $\eta p$, $K\Lambda$, $K\Sigma$, $\pi^+\pi^-p$, etc. The $N^*$ studies in both photo- and electroproduction offer complementary
information on the resonance spectrum and structure in comparison with the hadroproduction studies.

%%%%%%%%%%%%%%%%%%%%%%%%%%%%%%%%%%%%%%%%%%%%%%%%%%%%%%%%%%%%%%%%%%%%%%%%%%%%%%%%%%%%%%%%%%%%%%%%%%%%%%%%%%%%%%%%%%%%%%%%%%%%%%%%%%%%%%%%%%%%
%%%%%%%%%%%%%%%%%%%%%%%%%%%%%%%%%%%%%%%%%%%%%%%%%%%%%%%%%%%%%%%%%%%%%%%%%%%%%%%%%%%%%%%%%%%%%%%%%%%%%%%%%%%%%%%%%%%%%%%%%%%%%%%%%%%%%%%%%%%%

Studies of the electroproduction process allow us to explore the structure of nucleon resonances. The $N^*$ electroexcitation amplitudes 
extracted from these data as a function of $Q^2$ are directly related to the structure of the $N^*$ states. The electroexcitation of nucleon
resonances can be fully described by two transverse electrocouplings, $A_{1/2}(Q^2)$, $A_{3/2}(Q^2)$, and by the longitudinal electrocoupling
$S_{1/2}(Q^2)$. These electrocouplings are proportional to the helicity amplitudes that describe the transition between the initial state 
virtual photon-target proton and the final $N^*$ states for different helicities of the photon and proton in their center-of-mass (CM) 
frame. These quantities are unambiguously determined through the connection of the $N^*$ electromagnetic decay widths to the final states 
with transversely ($\Gamma_\gamma^T$) and longitudinally ($\Gamma_\gamma^L$) polarized photons at the resonant point ($W=M_r$) via:

\begin{equation}
\label{Eq:EMWidths1}
\Gamma_\gamma^T(W=M_r,Q^2)=\frac{q^2_{\gamma,r}(Q^2)}{\pi}\frac{2M_N}{(2J_r+1)M_r} \left( |A_{1/2}(Q^2)|^2+|A_{3/2}(Q^2)|^2 \right),
\end{equation}
\begin{equation}
\label{Eq:EMWidths2}
\Gamma_\gamma^L(W=M_r,Q^2)=2\frac{q^2_{\gamma,r}(Q^2)}{\pi}\frac{2M_N}{(2J_r+1)M_r}|S_{1/2}(Q^2)|^2,
\end{equation}
\noindent
with $q_{\gamma,r}=\left.q_{\gamma} \right|_{W=M_r}$. Here $M_r$ and $M_N$ represent the masses of the $N^*$ and ground state nucleon, 
respectively, and $J_r$ is the $N^*$ spin. In these expressions, the electrocouplings should have dimension [GeV$^{-1/2}$]. The transverse 
$A_{1/2}$ and $A_{3/2}$ couplings at the photon point ($Q^2$=0) fully describe nucleon resonance photoexcitation. The longitudinal $S_{1/2}$
coupling becomes irrelevant in photoproduction processes for which the flux of longitudinally polarized photons is zero, owing to gauge 
invariance of quantum electrodynamics (QED). Frequently, the absolute values of the $A_{1/2}$, $A_{3/2}$, and $S_{1/2}$ photo-/electrocouplings 
are presented after multiplication by 1000. More details on the nucleon resonance photo- and electrocouplings can be found in 
Ref.~\cite{azbu12}.

%%%%%%%%%%%%%%%%%%%%%%%%%%%%%%%%%%%%%%%%%%%%%%%%%%%%%%%%%%%%%%%%%%%%%%%%%%%%%%%%%%%%%%%%%%%%%%%%%%%%%%%%%%%%%%%%%%%%%%%%%%%%%%%%%%%%%%%%%%%%%%
\begin{figure*}
  \raisebox{6mm}{\includegraphics[width=0.64\textwidth,height=8.0cm]{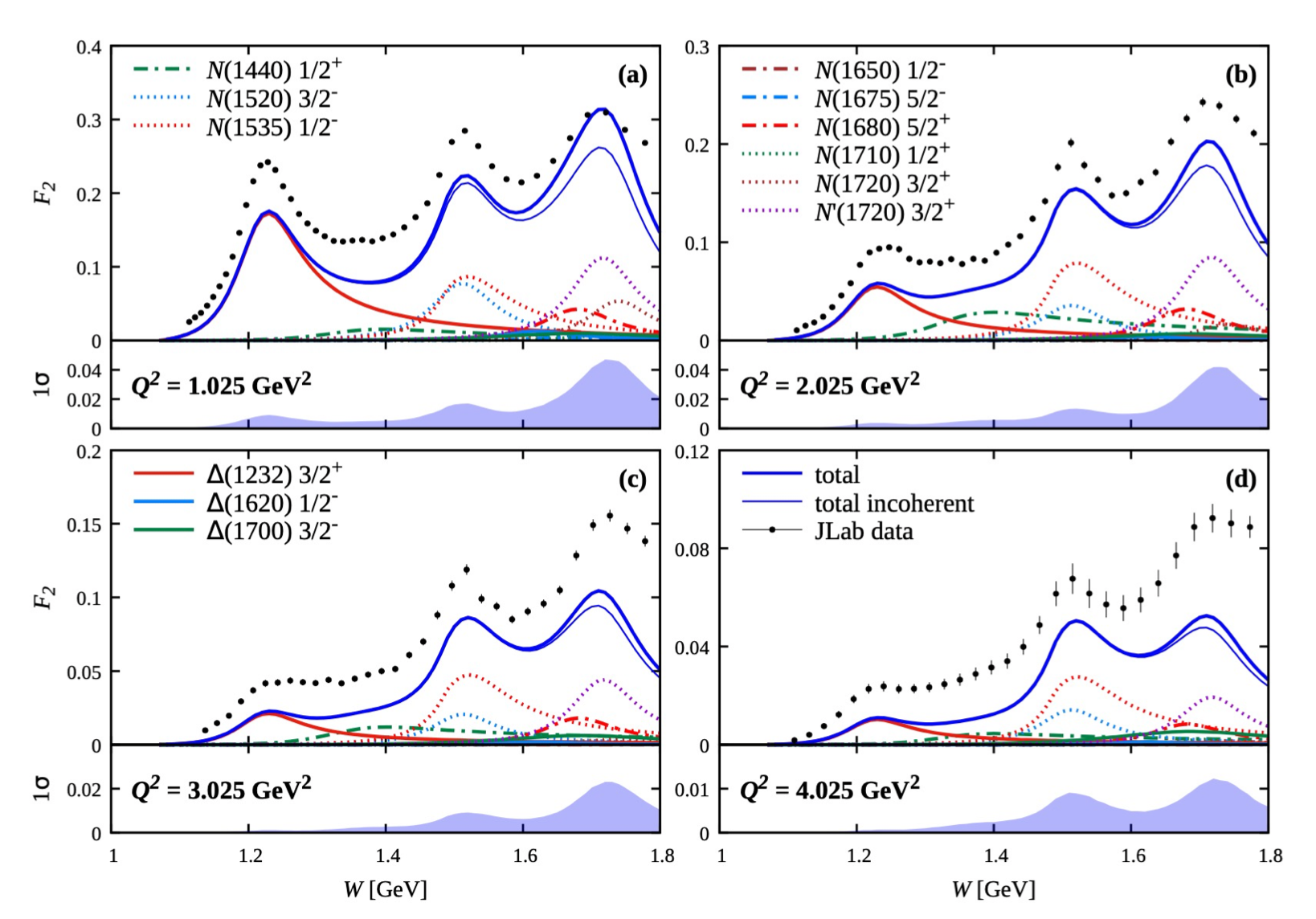}}
  \raisebox{0mm}{\includegraphics[width=0.35\textwidth]{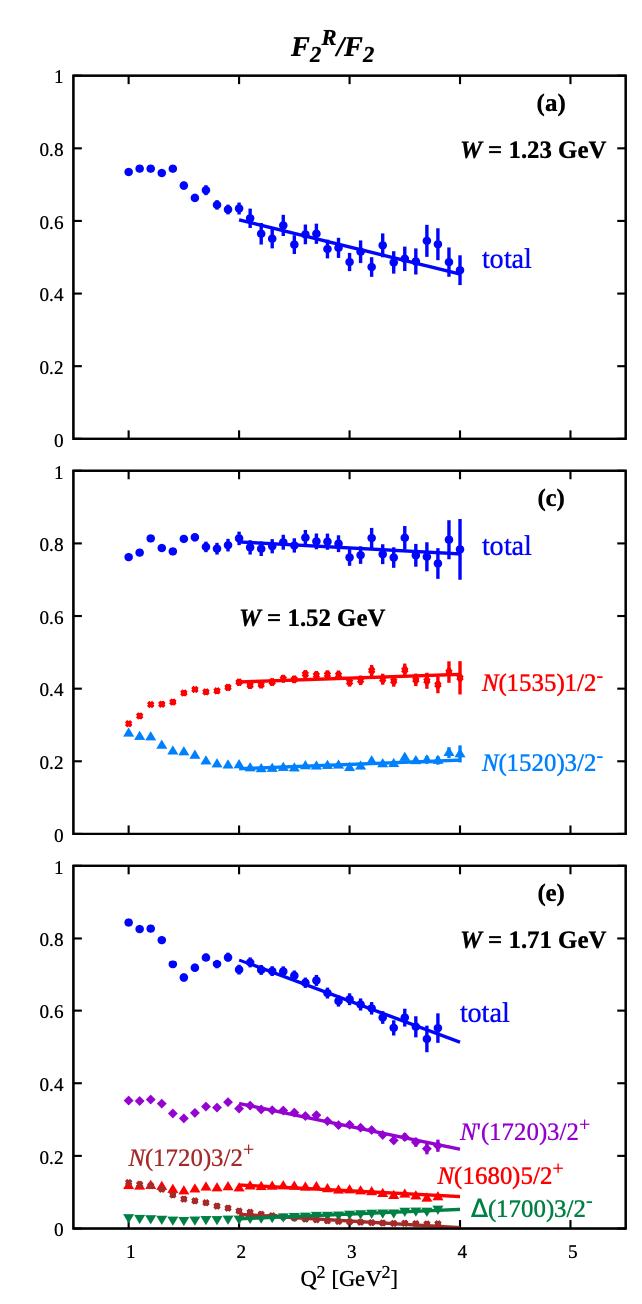}}
\caption{(Left) Resonant contributions to the inclusive $F_2$ structure function evaluated from the CLAS results on the $\gamma_vpN^*$
electrocouplings~\cite{blin2019,blin2021}. The resonant contributions shown are obtained as the coherent sum of the Breit-Wigner amplitudes
(thick, solid curves) and the incoherent sum of the cross sections from the individual resonances (thin, solid curves). (Right) The ratio of 
the $F_2$ resonant contributions to the full inclusive $F_2$ as a function of $Q^2$ at $W$-values corresponding to the peak locations in the 
first, second, and third resonance regions.}
\label{incl_res}
\end{figure*}
%%%%%%%%%%%%%%%%%%%%%%%%%%%%%%%%%%%%%%%%%%%%%%%%%%%%%%%%%%%%%%%%%%%%%%%%%%%%%%%%%%%%%%%%%%%%%%%%%%%%%%%%%%%%%%%%%%%%%%%%%%%%%%%%%%%%%%%%%%%%%%

Studies of meson electroproduction in the nucleon resonance region with the CLAS detector provided the first and only available results on
the electrocouplings of most nucleon resonances in the mass range up to 1.8~GeV~\cite{bur2020,fbs-carman,mok2020}. These results allow for 
the evaluation of the resonant contributions into inclusive electron scattering from the experimental data on the electrocouplings
\cite{blin2019,blin2021,blin2021a}. The evaluated resonant contributions into the inclusive $F_2$ structure function are shown in
Fig.~\ref{incl_res}. The peak in the first resonance region is mainly due to the contribution of the well-isolated $\Delta(1232)3/2^+$ 
resonance. The structures in the second and third resonance regions are generated by multiple overlapping resonances. The differences in the
structure of the contributing resonances in the three resonance regions result in pronounced differences in the $Q^2$-evolution of the 
resonant contributions into the inclusive cross section and structure functions seen in the three resonance regions. Therefore, studies 
of the electrocouplings of all prominent nucleon resonances are critical for gaining insight into the many facets of the dynamics of strong 
QCD responsible for the generation of the structure of the nucleon ground and excited states.

\section{Facilities for $N^*$ Studies in Exclusive Meson Photo- and Electroproduction}

The results shown in Fig.~\ref{incl_res} demonstrate that the resonance peaks seen in the second and third resonance regions are due to 
several overlapping states. In order to disentangle their individual contributions, it is important to explore the different exclusive 
meson photo- and electroproduction channels that have selective sensitivity to the different $N^*$s. For this purpose, the differences 
in the angular distributions of the decay products from the resonances of different spins and parities can be helpful. This consideration 
makes the measurements from detectors of nearly $4\pi$ acceptance of particular importance. These studies rely upon the exploration of all 
prominent exclusive photo- and electroproduction channels with electron beams of nearly 100\% duty factor. The beam duty factor $D$ is defined 
as the product of the pulse length and the repetition frequency. The counting rate for exclusive events $R_{exc}$ and the rate of accidental 
events $R_{acc}$ are related to the beam current $i_b$ as $R_{exc} \sim i_b$, $R_{acc} \sim (i_b/D)^{n_c}$, where $n_c$ is the number of 
measured final state particles in coincidence. Continuous beams of $\sim$100\% duty factor are essential in order to minimize the rate of 
accidental events, especially in the exploration of multi-meson exclusive channels. The facilities for $N^*$ studies with continuous electron 
beams and detectors of $\sim$4$\pi$ acceptance for studies of exclusive photo- and electroproduction are listed in Table~\ref{photo_fac}
\cite{ireland2020} and Table~\ref{electro_fac}~\cite{azbu12}, respectively.

%%%%%%%%%%%%%%%%%%%%%%%%%%%%%%%%%%%%%%%%%%%%%%%%%%%%%%%%%%%%%%%%%%%%%%%%%%%%%%%%%%%%%%%%%%%%%%%%%%%%%%%%%%%%%%%%%%%%%%%%%%%%%%%%%%%%%%%%%%%
\begin{table*}[htb]
\begin{center}
\vspace{2mm}
\begin{tabular}{|c|c|c|c|} \hline
Facility    & $W$ coverage, & Beam type/Flux &  Country      \\
            & GeV           &                        &      \\ \hline
CLAS/JLab   &  1.1-3.0      & Tagged bremsstrahlung  & USA  \\
            &               & 10$^8$ $\gamma$/s      & decommissioned \\ \hline
CLAS12/JLab &  1.1-4.0      & Quasi-real photons     & USA  \\
            &               & $L$=10$^{35}$ cm$^{-2}$s$^{-1}$ & operational \\ \hline  
Graal       &  1.1-1.9      & Compton scattering     & France \\
            &               & 10$^6$ $\gamma$/s      & decommissioned \\ \hline
MAMI        &  1.1-1.9      &  Tagged bremsstrahlung & Germany \\
            &               & 10$^7$ $\gamma$/s      & operational \\ \hline 
ELSA        &  1.1-2.7      &  Tagged bremsstrahlung & Germany \\
            &               & 10$^7$ $\gamma$/s      & operational \\ \hline    
SPring-8    &  1.1-2.5      & Compton scattering     & Japan \\
            &               & 10$^6$ $\gamma$/s      & operational \\ \hline
\end{tabular}
\end{center}
\caption{Facilities for nucleon resonance studies in exclusive photoproduction experiments~\cite{ireland2020}. For the CLAS12 experiments 
with quasi-real photons, the luminosity ${\cal L}$ for reactions induced by the electron beam on the liquid-hydrogen target~\cite{clas12-nim} 
is presented.}
\label{photo_fac}
\end{table*}
%%%%%%%%%%%%%%%%%%%%%%%%%%%%%%%%%%%%%%%%%%%%%%%%%%%%%%%%%%%%%%%%%%%%%%%%%%%%%%%%%%%%%%%%%%%%%%%%%%%%%%%%%%%%%%%%%%%%%%%%%%%%%%%%%%%%%%%%%%%%

%%%%%%%%%%%%%%%%%%%%%%%%%%%%%%%%%%%%%%%%%%%%%%%%%%%%%%%%%%%%%%%%%%%%%%%%%%%%%%%%%%%%%%%%%%%%%%%%%%%%%%%%%%%%%%%%%%%%%%%%%%%%%%%%%%%%%%%%%%%
\begin{table*}[htb]
\begin{center}
\vspace{2mm}
\begin{tabular}{|c|c|c|c|c|} \hline
Facility    & $Q^2$ coverage, & $W$ coverage, & Luminosity,        & Country      \\
            & GeV$^2$         & GeV           & cm$^{-2}$ s$^{-1}$ &          \\ \hline
CLAS/JLab   &  0.2-5.0        & 1.1-2.5       & 10$^{34}$          & USA  \\
            &                 &               &                    & decommissioned \\ \hline
CLAS12/JLab &  0.05-10.0      & 1.1-4.0       & 10$^{35}$          & USA  \\
            &                 &               &                    & operational  \\ \hline
Hall~C/JLab &  $<$ 7.0        & 1.1-2.8       & 10$^{37}$          & USA  \\
            &                 &               &                    & operational  \\ \hline
MAMI        &  $<$ 0.2        & 1.1-1.3       & 10$^{36}$          & Germany  \\
            &                 &               &                    & operational  \\ \hline     
MIT/Bates   &  $<$ 0.15       & 1.1-1.3       & 10$^{36}$          & USA  \\
            &                 &               &                    & decommissioned  \\ \hline
\end{tabular}
\end{center}
\caption{Facilities for nucleon resonance studies in exclusive electroproduction experiments~\cite{azbu12,clas12-nim}.}
\label{electro_fac}
\end{table*}
%%%%%%%%%%%%%%%%%%%%%%%%%%%%%%%%%%%%%%%%%%%%%%%%%%%%%%%%%%%%%%%%%%%%%%%%%%%%%%%%%%%%%%%%%%%%%%%%%%%%%%%%%%%%%%%%%%%%%%%%%%%%%%%%%%%%%%%%%%%%

The unique combination of multi-GeV continuous electron beams and a detector of $\sim$4$\pi$ acceptance made the CLAS facility in Hall~B at 
JLab~\cite{mecking2003} the best for studies of the spectrum and structure of $N^*$ states from exclusive meson photo- and electroproduction 
data. CLAS allowed for the study of exclusive electroproduction reactions in the range of $Q^2$ up to 5~GeV$^2$ and $W$ up to 3~GeV. These 
data have provided the dominant part of the available world information on the $\pi N$, $\eta p$, $K \Lambda$, $K \Sigma$, and $\pi^+\pi^- p$
electroproduction channels in the resonance region ($W < 2.5$~GeV) with almost complete coverage of the final state CM phase space. The 
majority of the experiments were carried out with an unpolarized liquid-hydrogen target and a longitudinally polarized electron beam. 
Approximately 200k data points for differential cross sections, separated structure functions, and single- and double-polarization observables 
have become available based on analyses of these experimental data, which are stored in the CLAS Physics Database~\cite{clasdb}. $N^*$ studies 
in exclusive meson electroproduction at $Q^2 < 7.5$~GeV$^2$ were carried out with the spectrometers of small millisteradian acceptance in 
Hall~C at JLab, but with luminosity two orders of magnitude larger than that achieved with CLAS. Nucleon resonance electroexcitation at small
$Q^2$ was explored at MAMI and MIT/Bates with spectrometers of small acceptance (see Table~\ref{electro_fac}). 

In the period from 2012 to 2017, CLAS was replaced with the new large acceptance CLAS12 spectrometer~\cite{clas12-nim} as part of the JLab 
12-GeV upgrade project. The extended program includes a number of experiments as part of the continuing $N^*$ program in Hall~B, which will 
collect data over an unprecedented kinematic range for the study of nucleon excited states in the range of $Q^2$ from 0.05~GeV$^2$ up to at 
least 10~GeV$^2$, spanning the full CM angular range of the decay final states.

The CLAS detector was also optimized for the studies of exclusive photoproduction with charged mesons in the final states. Unpolarized and 
circularly polarized photons produced by passage of the electron beam through a thin gold radiator were used in photoproduction experiments, 
along with linearly polarized photons from coherent bremsstrahlung on a diamond radiator. The energy of the bremsstrahlung photons was tagged 
by measuring the recoil electron in the Hall~B tagging spectrometer~\cite{sober2000} located upstream of CLAS. Two different frozen-spin 
polarized targets were used in photoproduction experiments, allowing both longitudinal and transverse proton polarizations~\cite{keith2012} 
and longitudinal proton and deuteron polarizations~\cite{lowry2016}.

Studies of exclusive meson photoproduction were also carried out at European and Asian facilities that combine continuous electron beams and
detectors of nearly $4\pi$ acceptance as listed in Table~\ref{photo_fac}. These facilities are optimized for the measurements of final 
states with neutral mesons and provide complementary information on nucleon resonances to that available from the JLab experiments. 

\section{Nucleon Resonance Parameters from Exclusive Meson Photoproduction}

The studies of exclusive meson photoproduction in experiments with continuous electron beams and large-acceptance detectors allow us to 
determine the types of all final state particles and their four-momenta in each reaction. Most exclusive photoproduction channels in the 
resonance region were studied~\cite{ireland2020,crede2013,klempt2010,krusche2003} and include results on differential cross sections and 
polarization asymmetries for different combinations/orientations of the photon beam, target nucleon, and the final state hadron polarization 
vectors.

Representative examples of the fully integrated cross sections for the major exclusive meson photoproduction channels in the resonance
region are shown in Fig.~\ref{integ_ex_photo}, together with the inclusive $(\gamma,X)$ cross section~\cite{beck2017}. The different 
exclusive channels clearly demonstrate selective sensitivities to different resonance contributions. The exclusive $\pi^0 p$ channel shows
sensitivity to the resonances located in the first, second, and third resonance regions. The pronounced structure at $W \sim 1.53$~GeV seen 
in the $\eta p$ cross section is related to the contribution from the $N(1535)1/2^-$ resonance. The $\pi\pi N$ channels are suitable for the
exploration of high-lying resonances in the mass range of $W > 1.6$~GeV, together with the $K\Lambda$ and $K\Sigma$ channels. The exclusive 
photoproduction channel amplitudes can be described by the sum of the resonant contributions and the complex set of non-resonant mechanisms 
(see Fig.~\ref{integ_ex_photo} top right panel). There are substantial differences in the non-resonant amplitudes in different channels. 
Instead, the $N^*$ photocouplings seen in all channels should be the same since the nucleon resonance photoexcitation and hadronic decay 
amplitudes are independent. Therefore, consistent results on the nucleon resonance photocouplings available from independent studies of 
different meson photoproduction channels validate the credible extraction of these quantities. 

The nucleon resonance parameters such as the masses, the total and partial hadronic decay widths, and the photocouplings, can be determined 
either within the framework of reaction models~\cite{drechsel2007,tiator2018,tiator2018a,aznauryan2003,aznauryan2003a,gol2019} or from the 
amplitudes directly determined from the measured observables of meson photoproduction data within so-called ``complete" measurements 
\cite{svarc2018,workman2017,workman2012}. In reaction models, the resonant amplitudes are frequently parameterized within the framework of
the Breit-Wigner ansatz. The non-resonant contributions are described by partial waves and/or by effective meson-baryon mechanisms described 
by the tree-level diagrams. In the different analyses, additional constraints from dispersion relations were employed for the photoproduction
amplitudes~\cite{drechsel2007,tiator2018,tiator2018a,aznauryan2003,aznauryan2003a}. The $N^*$ parameters were determined from fits to all 
available measured observables of the exclusive reaction channels under study.

The impressive progress achieved in the last decade with the measurements of the photoproduction differential cross sections and polarization
asymmetries~\cite{bur2020,ireland2020,crede2013} has provided a dataset for ``complete" measurements. Combined analyses of these observables 
makes it possible to determine the pseudoscalar meson-baryon photoproduction amplitudes for the $\pi N$, $K\Lambda$, and $K\Sigma$ channels 
directly from the experimental data. These reactions can be described by eight independent amplitudes as a function of the Mandelstam variables 
$s$ and $t$ (or alternatively, $W$ and the polar angle $\theta_m$ of the final state meson in the CM frame) and the azimuthal angle $\phi_m$ 
between the reaction plane and the plane composed by the target polarization vector and the beam photon momentum. The number of independent 
amplitudes can be obtained by counting the number of helicity states for the initial and final state particles. Both the beam photon and the 
target proton have two possible helicity quantum numbers. The final state baryon of spin 1/2 also has two helicity quantum numbers, while 
the pseudoscalar meson exists in a single helicity 0 state. Overall, $2\times2\times2=8$ helicity amplitudes describe pseudoscalar meson-baryon
photoproduction. Parity conservation reduces this number down to four. The amplitudes are complex numbers that have real and imaginary parts,
hence, eight real numbers fully describe these photoproduction amplitudes. Therefore, a minimum of eight well chosen independent observables 
are needed in order to determine these amplitudes from the data. 

%%%%%%%%%%%%%%%%%%%%%%%%%%%%%%%%%%%%%%%%%%%%%%%%%%%%%%%%%%%%%%%%%%%%%%%%%%%%%%%%%%%%%%%%%%%%%%%%%%%%%%%%%%%%%%%%%%%%%%%%%%%%%%%%%%%%%%%%%%%%
\begin{figure}
  \includegraphics[width=0.35\textwidth,height=3.3cm]{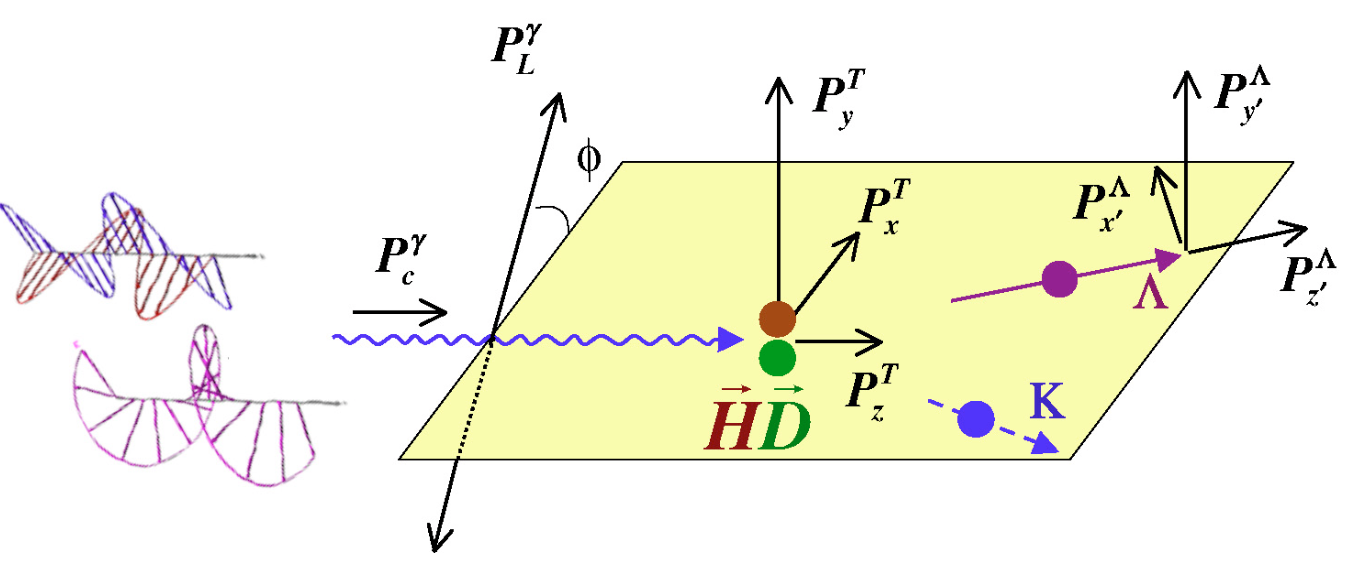}
  \raisebox{0mm}{\includegraphics[width=0.65\textwidth]{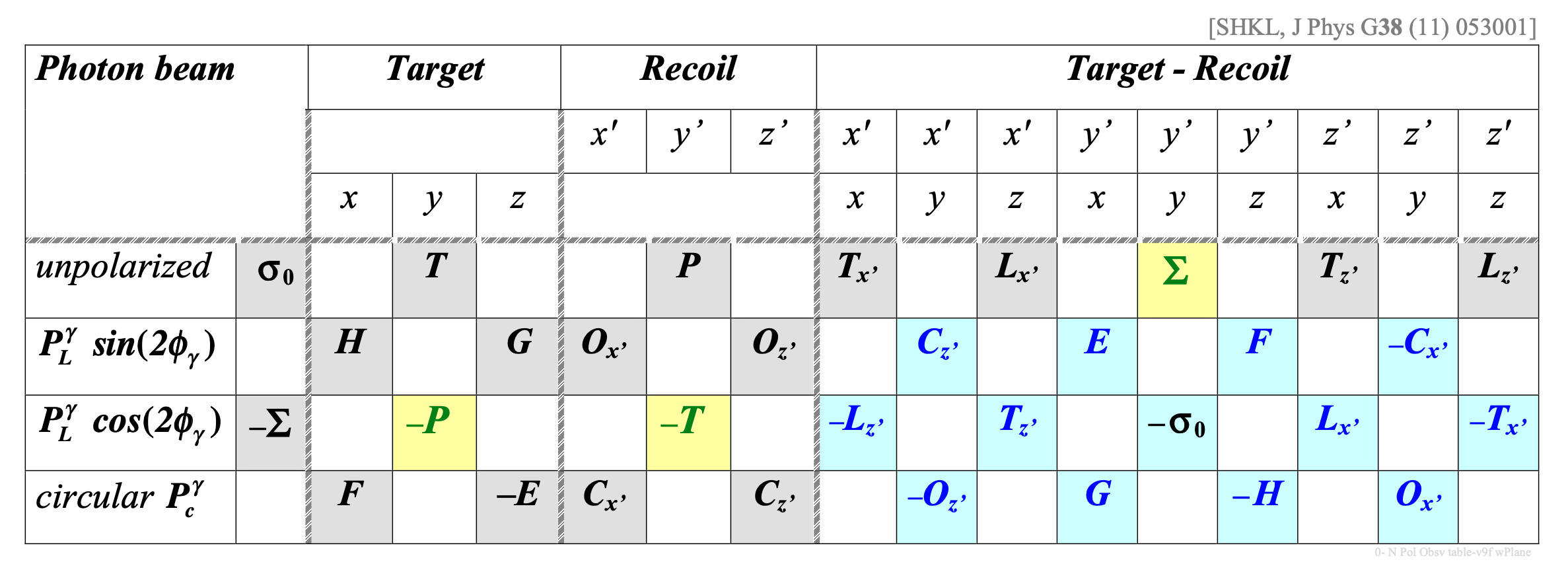}}
\caption{(Left) The initial and final state particle polarizations in the measurements of polarization observables in pseudoscalar 
meson-baryon photoproduction. (Right) Polarization observables accessible from pseudoscalar meson-baryon photoproduction in experiments with
a polarized photon beam, proton target, and measurements of the recoil hyperon polarization~\cite{sandorfi2011}.}
\label{pol_observables}
\end{figure}
%%%%%%%%%%%%%%%%%%%%%%%%%%%%%%%%%%%%%%%%%%%%%%%%%%%%%%%%%%%%%%%%%%%%%%%%%%%%%%%%%%%%%%%%%%%%%%%%%%%%%%%%%%%%%%%%%%%%%%%%%%%%%%%%%%%%%%%%%%%%

The observables of pseudoscalar meson-baryon photoproduction available from experiments with polarized photon beam, proton target, and 
from measurements of the recoil hyperon polarization are detailed in Fig.~\ref{pol_observables}. They can be inferred from the angular 
distributions of the final state meson over the azimuthal angle $\phi_m$ from the measurements with different orientations of the photon 
and proton polarization vectors, or with circularly polarized photons. In addition, in the $K\Lambda$ and $K\Sigma$ channels, the hyperon
polarization can be obtained from the angular distributions of the hyperon decay products, providing access to double and triple polarization 
observables.

\begin{figure*}
\begin{center}
  \raisebox{0mm}{\includegraphics[width=0.98\textwidth,height=7.0cm]{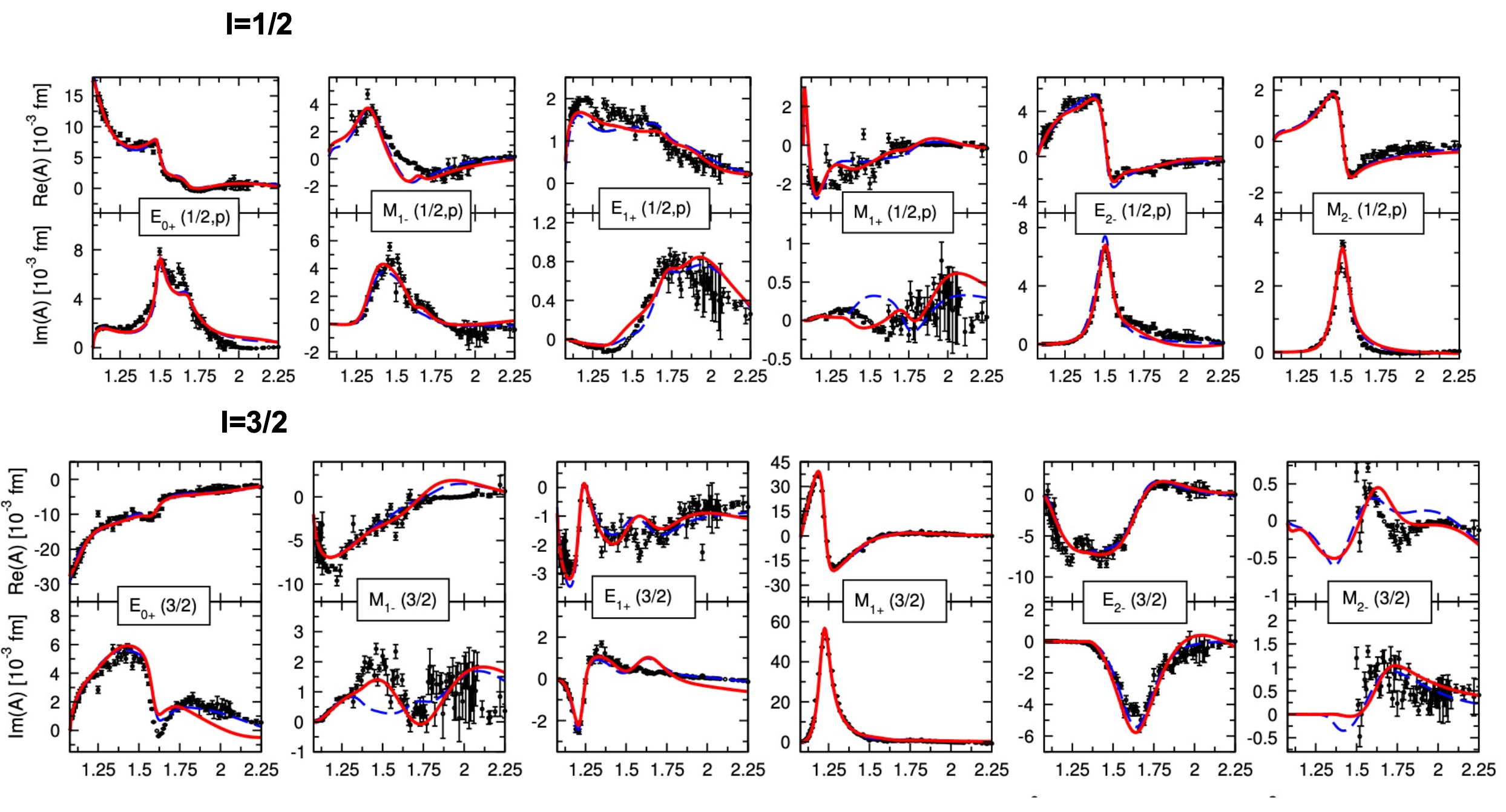}}
\vspace{-2mm}
\caption{$\pi N$ photoproduction amplitudes extracted directly from the complete measurements of the unpolarized cross sections and 
polarization observables after decomposition over the partial waves~\cite{workman2017}. The results for the final states of isospin 1/2 and 
3/2 are shown in the top and bottom panels, respectively. The real and imaginary parts of the multipoles are shown in the top and bottom 
rows of each panel of plots, respectively.}
\label{ampl_pw}
\end{center}
\end{figure*}

The angular distribution of the final state meson in the CM frame over the azimuthal angle $\phi_m$ is given by Eqs.(1a-p) in 
Ref.~\cite{ireland2020} in terms of the polarization observables listed in Fig.~\ref{pol_observables}, while the connection between these 
observables and the well-known Chew-Goldberger-Low-Nambu (CGLN) amplitudes is given by Eqs.(58a-p) in Ref.~\cite{sandorfi2011}. The relations
between the amplitudes in different representations can be found in Ref.~\cite{knochlein1995}. As of now, eight of the observables listed in
Fig.~\ref{pol_observables} from $\pi N$ photoproduction data and sixteen observables from $K\Lambda$ photoproduction data have become
available~\cite{ireland2020}. They allow for the extraction of the pseudoscalar meson-baryon amplitudes directly from the experimental data,
avoiding reaction model bias by fitting the amplitudes to all the observables. The results for the extracted $\pi N$ photoproduction 
amplitudes are shown in Fig.~\ref{ampl_pw} after decomposition over the partial waves. Each partial wave corresponds to the production 
amplitudes for the final state of isospin $I$, orbital angular momentum $L$ shown by the subscript index of the multipole, and the total 
angular momentum $J$, which is equal to either $L+1/2$ or $L-1/2$ for plus or minus signs in the multipole subscript index, respectively.

%%%%%%%%%%%%%%%%%%%%%%%%%%%%%%%%%%%%%%%%%%%%%%%%%%%%%%%%%%%%%%%%%%%%%%%%%%%%%%%%%%%%%%%%%%%%%%%%%%%%%%%%%%%%%%%%%%%%%%%%%%%%%%%%%%%%%%%%%%%%
%%%%%%%%%%%%%%%%%%%%%%%%%%%%%%%%%%%%%%%%%%%%%%%%%%%%%%%%%%%%%%%%%%%%%%%%%%%%%%%%%%%%%%%%%%%%%%%%%%%%%%%%%%%%%%%%%%%%%%%%%%%%%%%%%%%%%%%%%%%%
 
For the extraction of the $N^*$ parameters from the amplitudes determined directly from the data, they should be analytically continued into 
the complex energy plane. The real and imaginary parts for the pole position in the second Riemann sheet can be related to the $N^*$ mass and 
total decay width, respectively. The residues at the pole position can be related to the product of the amplitudes for the resonance 
photoexcitation and decay to the final state under study~\cite{kamano2016}. The data from a complete measurement paves the way to a nearly 
model-independent extraction of the $N^*$ spectrum and photocouplings from exclusive photoproduction data. 

In the extraction of the $N^*$ parameters from the exclusive meson photo- and electroproduction data, it is important to account for final 
state interactions (FSIs) as schematically depicted in Fig.~\ref{integ_ex_photo} (bottom right). In these processes many intermediate $M'B'$ 
and $M'M''B'$ states can be populated in the kinematically allowed channels, and through a sequence of hadron re-scattering, they can populate 
the $MB$ final state under study. Also, the amplitude for the production of the $MB$ final state under study can be reduced owing to its 
hadronic interaction with other meson-baryon channels. Several coupled-channel approaches have been developed allowing for the combined 
extraction of $N^*$ parameters from global multi-channel analyses of all available meson hadro- and photoproduction data. They include the
approaches by the Bonn-Gatchina~\cite{anisovich2012,anisovich2017}, Argonne-Osaka~\cite{kamano2016,kamano2013}, and
J{\"u}lich-Bonn-Washington~\cite{roenchen2014,roenchen2019} groups. These approaches have provided information on the $N^*$ photocouplings and 
their hadronic decay widths, allowing for a rigorous accounting of the constraints imposed on the production amplitudes by the general 
unitarity condition. The red lines in Fig.~\ref{ampl_pw} represent the $\pi N$ photoproduction amplitudes evaluated from a global multi-channel
analysis~\cite{roenchen2014}. They are in good agreement with the amplitudes extracted directly from the complete $\pi N$ photoproduction
data~\cite{workman2012}.

\section{The Search for ``Missing" Resonances}

\begin{figure*}
\begin{center}
  \raisebox{0mm}{\includegraphics[width=0.80\textwidth,height=7.0cm]{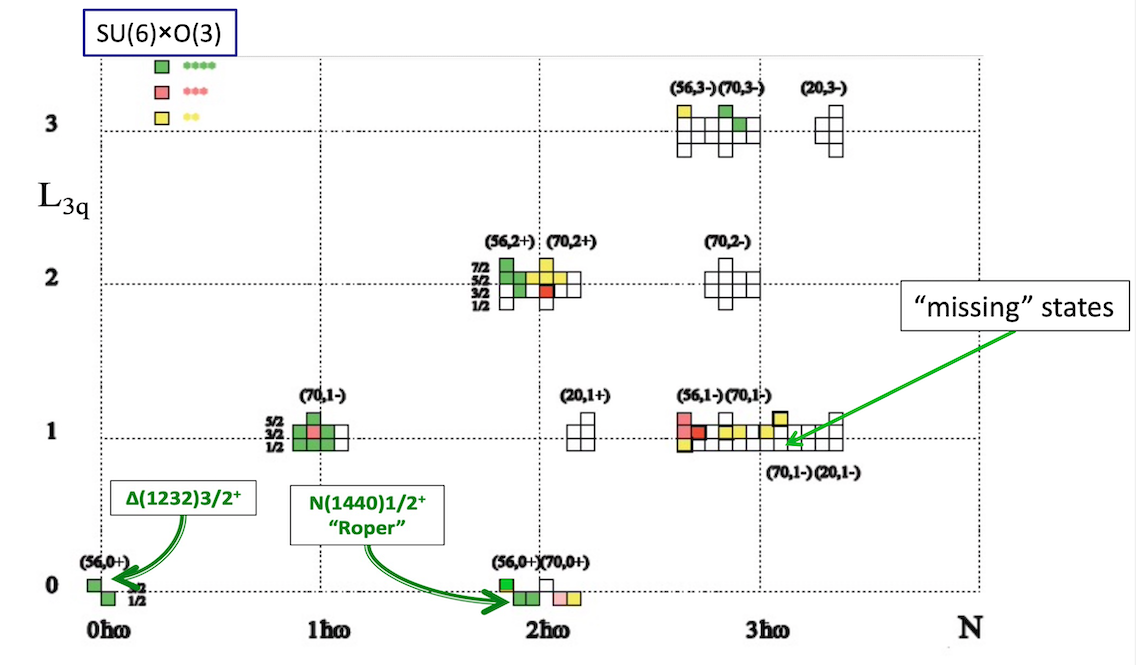}}
\vspace{-2mm}
\caption{Spectrum of nucleon resonances expected in quark models employing ${\rm SU(6)}_{\rm spin-flavor} \times {\rm O(3)}_{\rm space}$ 
symmetry. $L_{3q}$ is the orbital angular momentum of the three constituent quarks and the quantum number $N$ corresponds to the radial 
excitation of the three-quark system. The predicted and observed states are shown by the filled boxes, while the predicted and still not 
observed states are shown by the open boxes.}
\label{qm_prediction}
\end{center}
\end{figure*}

Constituent quark models, based on approximate symmetries of the strong interaction Hamiltonian established from the experimental results on 
the $N^*$ spectrum known before 2012~\cite{klempt2010,capstick2000,capstick1986,giannini2015}, predict many more excited states of the
nucleon than have been established from experiment. The expectation from quark models that employ SU(6)$\times$O(3)
(spin-flavor$\times$space-rotational) symmetry is depicted in Fig.~\ref{qm_prediction}. The predicted and observed nucleon resonances are 
shown by the filled boxes. The states that are predicted but still not observed are shown by the open boxes. These expectations gain support 
from the $N^*$ spectrum studies starting from the QCD Lagrangian both within lattice QCD~\cite{edwards2011} and the continuum Schwinger
method approach (CSM)~\cite{chen2019a,qin2019}. The search for states in the mass range above 1.7~GeV that have eluded detection has become 
the focus of extensive studies to address the so-called ``missing" resonance problem. 

%%%%%%%%%%%%%%%%%%%%%%%%%%%%%%%%%%%%%%%%%%%%%%%%%%%%%%%%%%%%%%%%%%%%%%%%%%%%%%%%%%%%%%%%%%%%%%%%%%%%%%%%%%%%%%%%%%%%%%%%%%%%%%%%%%%%%%%%%%%%
%%%%%%%%%%%%%%%%%%%%%%%%%%%%%%%%%%%%%%%%%%%%%%%%%%%%%%%%%%%%%%%%%%%%%%%%%%%%%%%%%%%%%%%%%%%%%%%%%%%%%%%%%%%%%%%%%%%%%%%%%%%%%%%%%%%%%%%%%%%%

%%%%%%%%%%%%%%%%%%%%%%%%%%%%%%%%%%%%%%%%%%%%%%%%%%%%%%%%%%%%%%%%%%%%%%%%%%%%%%%%%%%%%%%%%%%%%%%%%%%%%%%%%%%%%%%%%%%%%%%%%%%%%%%%%%%%%%%%%%%%
\begin{figure*}
\begin{center}
  \raisebox{0mm}{\includegraphics[width=0.98\textwidth,height=4.5cm]{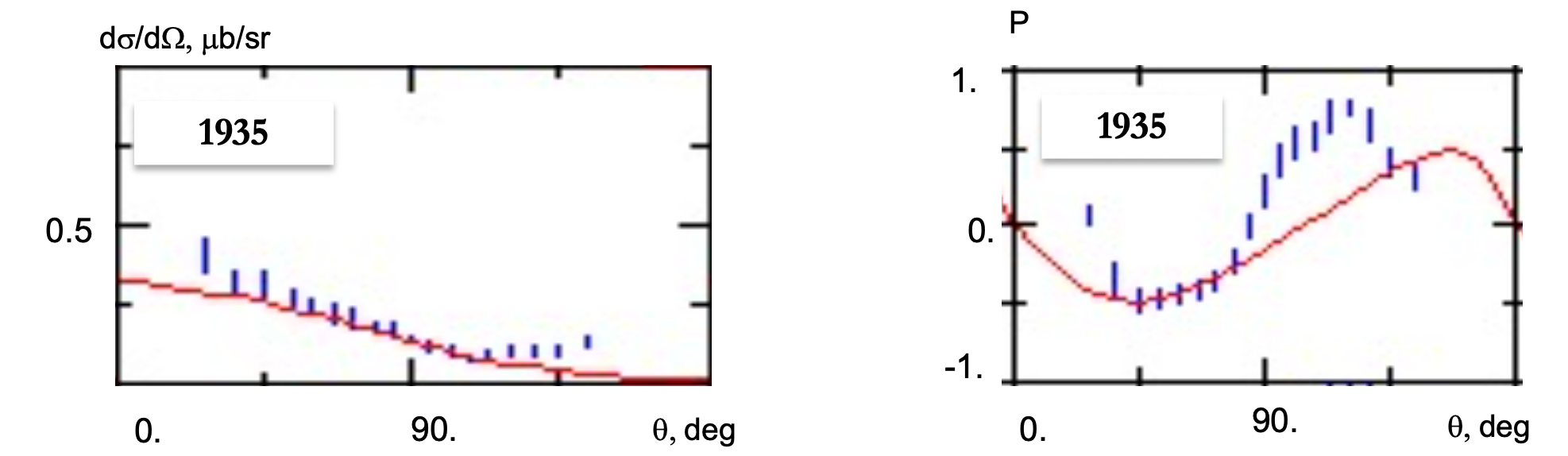}}
  \raisebox{0mm}{\includegraphics[width=0.95\textwidth,height=4.5cm]{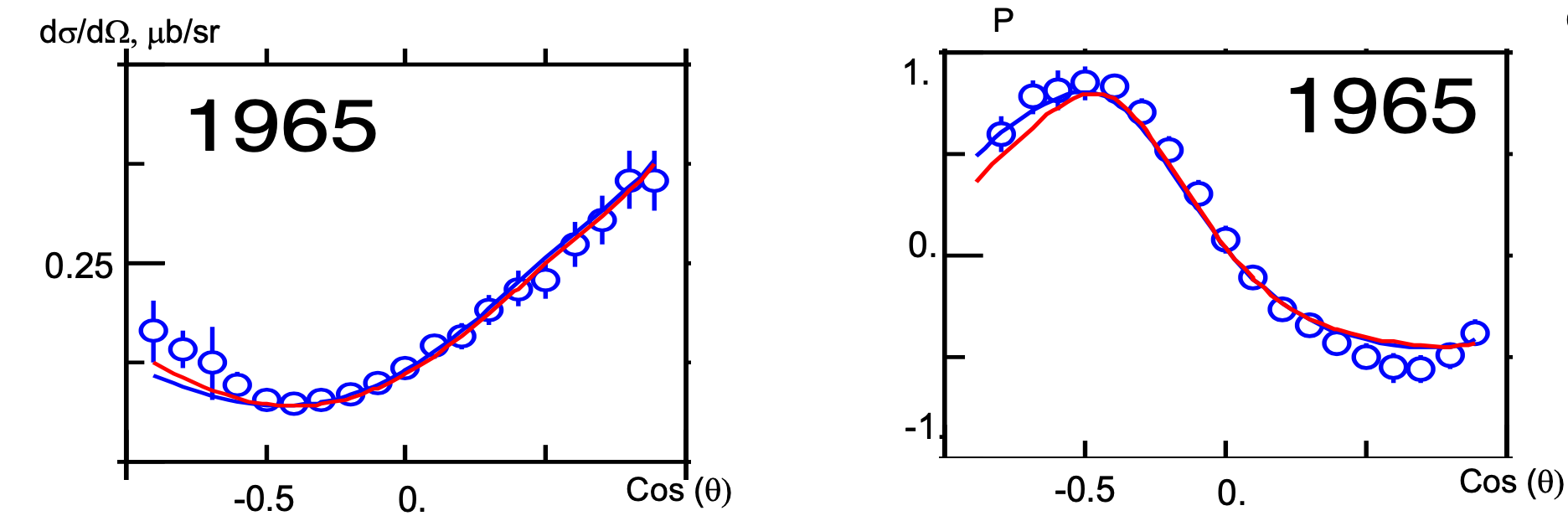}}
\vspace{0mm}
\caption{Representative examples for the description of the CLAS results on the $K\Lambda$ differential cross sections (left column) and 
induced polarization (right column)~\cite{mcc2009} achieved within the Argonne-Osaka coupled-channel approach (red lines)~\cite{kamano2013} 
with the well-established resonances before 2012 (top row) and the Bonn-Gatchina coupled-channel approach (red lines)~\cite{anisovich2012} 
after implementation of nine new nucleon resonances (bottom row). The $W$ value in MeV is given in the top corner of each plot.}
\label{KL_aobg}
\end{center}
\end{figure*}
%%%%%%%%%%%%%%%%%%%%%%%%%%%%%%%%%%%%%%%%%%%%%%%%%%%%%%%%%%%%%%%%%%%%%%%%%%%%%%%%%%%%%%%%%%%%%%%%%%%%%%%%%%%%%%%%%%%%%%%%%%%%%%%%%%%%%%%%%%%%

Recently, several long-awaited new nucleon resonances were discovered in global multi-channel analyses of exclusive meson photo- and
hadroproduction data~\cite{anis2017,ron2014} with a decisive impact of the CLAS results on $K\Lambda$ and $K\Sigma$ photoproduction
\cite{brad2007,mcc2009,dey2010}. The CLAS results on the differential cross sections and induced polarization $P$ for $K\Lambda$ 
photoproduction, along with their description within coupled-channel approaches developed by the Argonne-Osaka~\cite{kamano2013} and 
Bonn-Gatchina~\cite{anisovich2012} groups, are shown in Fig.~\ref{KL_aobg}. Both approaches incorporate most exclusive $\pi N$ 
and photoproduction channels relevant in the resonance region. There are substantial differences in accounting for the resonant contributions 
in these two approaches. The Argonne-Osaka group includes all resonances known before 2012 with three- or four-star PDG status and, in general, 
the measured observables included are well reproduced in the kinematic ranges covered by the data, except for the interval 
1.8~GeV $< W < 2.0$~GeV. Here, the data on the $K\Lambda$ differential cross sections and induced polarization $P$ are not well reproduced, as 
can be seen in Fig.~\ref{KL_aobg} (top). The observed discrepancies emphasize the importance of the polarization data and demonstrate that 
accounting for the contributions from only the conventional resonances does not allow for the description of the data. The Bonn-Gatchina
coupled-channel approach accounts for the contributions from all well-established resonances and from nine new resonances with masses, hadronic
decay widths, and photocouplings fit to data from the meson hadro- and photoproduction channels. After implementation of the new baryon states, 
a good description of the CLAS data is achieved over the entire kinematic range as shown in Fig.~\ref{KL_aobg} (bottom). Furthermore, the
Bonn-Gatchina approach provides a good description of the observables for most relevant exclusive meson hadro- and photoproduction channels in 
the nucleon resonance region after implementation of the new states. This provides strong evidence for the existence of these new nucleon
resonances.

%%%%%%%%%%%%%%%%%%%%%%%%%%%%%%%%%%%%%%%%%%%%%%%%%%%%%%%%%%%%%%%%%%%%%%%%%%%%%%%%%%%%%%%%%%%%%%%%%%%%%%%%%%%%%%%%%%%%%%%%%%%%%%%%%%%%%%%%%%%%
\begin{figure*}
\begin{center}
  \raisebox{0mm}{\includegraphics[width=0.98\textwidth,height=7.0cm]{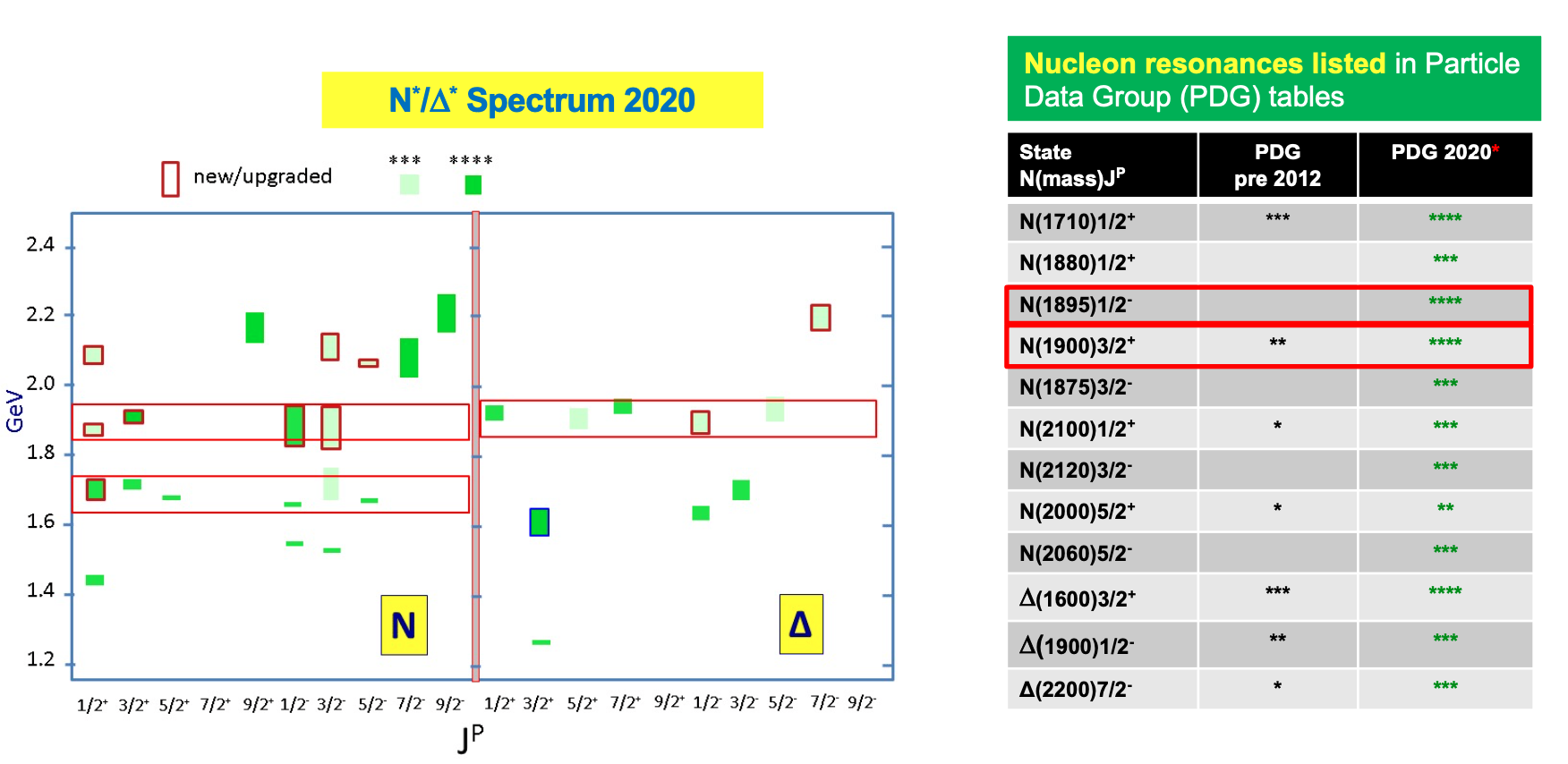}}
\vspace{-2mm}
\caption{(Left) Spectrum of the excited states of the nucleon established in 2020~\cite{bur2020}. (Right) The nucleon resonance PDG status 
before 2012 and in 2020~\cite{pdg}.}
\label{spectrum}
\end{center}
\end{figure*}
%%%%%%%%%%%%%%%%%%%%%%%%%%%%%%%%%%%%%%%%%%%%%%%%%%%%%%%%%%%%%%%%%%%%%%%%%%%%%%%%%%%%%%%%%%%%%%%%%%%%%%%%%%%%%%%%%%%%%%%%%%%%%%%%%%%%%%%%%%%%

The spectrum of nucleon resonances, as it is now known, is shown in Fig.~\ref{spectrum} (left) with the new states highlighted by the brown
boxes. Figure~\ref{spectrum} (right) shows a comparison between the PDG status of nucleon resonances in the mass range $W > 1.6$~GeV in 2012 
and 2020. Two of the nine recently discovered resonances get the highest four-star status of firmly established states. The status of the
other new states was elevated to three stars as likely existing. Discovery of the long-awaited new nucleon resonances with decisive impact 
from the CLAS $K\Lambda$ and $K\Sigma$ photoproduction data, represents an important achievement in hadron physics in the past decade.

%%%%%%%%%%%%%%%%%%%%%%%%%%%%%%%%%%%%%%%%%%%%%%%%%%%%%%%%%%%%%%%%%%%%%%%%%%%%%%%%%%%%%%%%%%%%%%%%%%%%%%%%%%%%%%%%%%%%%%%%%%%%%%%%%%%%%%%%%%%%
\begin{figure*}
\begin{center}
  \raisebox{0mm}{\includegraphics[width=0.98\textwidth,height=5.0cm]{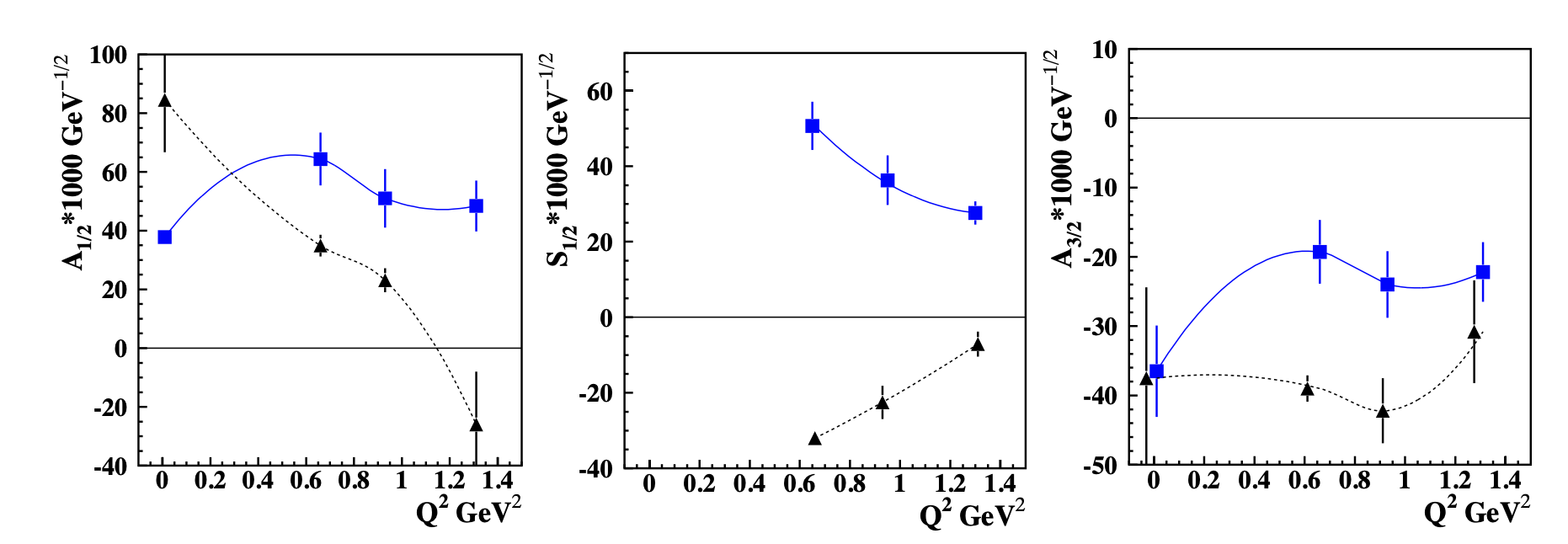}}
\vspace{-2mm}
\caption{Photo- and electroexcitation amplitudes of the new $N'(1720)3/2^+$ resonance (blue points connected by blue solid lines) and the
conventional $N(1720)3/2^+$ (black points connected by black dotted lines) obtained in Ref.~\cite{mokeev2020a} from the combined analysis of
$\pi^+\pi^-p$ photo- and electroproduction data.}
\label{miss_conv_electr}
\end{center}
\end{figure*}
%%%%%%%%%%%%%%%%%%%%%%%%%%%%%%%%%%%%%%%%%%%%%%%%%%%%%%%%%%%%%%%%%%%%%%%%%%%%%%%%%%%%%%%%%%%%%%%%%%%%%%%%%%%%%%%%%%%%%%%%%%%%%%%%%%%%%%%%%%%%

Combined studies of exclusive meson photo- and electroproduction data represent the next step in the search for new excited states of the 
nucleon. New resonances seen in photoproduction can be observed in electroproduction data with $Q^2$-independent values of their masses and 
hadronic decay widths. These studies, therefore, will validate the existence of these new states in a nearly model-independent way.

The new $N'(1720)3/2^+$ state has recently been discovered in the combined studies of $\pi^+\pi^-p$ photo- and electroproduction data measured 
with CLAS~\cite{mokeev2020a}, in addition to the nine new resonances already discussed. Implementation of the $N'(1720)3/2^+$ is required in 
order to describe the $\pi^+\pi^-p$ reaction in the third resonance region in the range 0.5~GeV$^2 < Q^2 < 1.5$~GeV$^2$ with $Q^2$-independent
masses and hadronic decay widths of the resonances relevant in this kinematic area. As of now, the $N'(1720)3/2^+$ is the only new resonance 
for which the $Q^2$-evolution of the electroexcitation amplitudes has been determined, as shown in Fig.~\ref{miss_conv_electr} in comparison 
with the conventional $N(1720)3/2^+$ state. These results offer an opportunity to gain insight into the structural features of the states that 
have made their observation so elusive. The structure of the new $N'(1720)3/2^+$ resonance has already been studied in analysis of its
electroexcitation amplitudes within the light-front holography approach~\cite{lub2020}.

\section{Nucleon Resonance Electroexcitation Amplitudes from Exclusive Meson Electroproduction}

The available and foreseen data from CLAS and CLAS12 on exclusive meson electroproduction in the resonance region open up a unique 
opportunity for the systematic exploration of the structure of all prominent nucleon resonances. The continuous, multi-GeV electron beam from 
the CEBAF accelerator, combined with a detector of nearly $4\pi$ acceptance, provides the best opportunity for the exploration of nucleon 
resonance structure from the results on the evolution of the nucleon resonance electroexcitation amplitudes ($\gamma_vpN^*$ electrocouplings) 
as a function of $Q^2$. The extensive research efforts in progress in Hall~B at JLab are focused on the extraction of the amplitudes of all 
prominent nucleon resonances in the mass range $W < 2.5$~GeV and 0.05~GeV$^2 < Q^2 < 10$~GeV$^2$ from both the independent and combined studies 
of the different meson electroproduction channels~\cite{Brodsky:2020vco,Burkert:2019bhp,bur2020,fbs-carman,azbu12}. Currently, the new CLAS12
detector, installed in Hall~B as the part of the JLab 12-GeV upgrade, is the only available and foreseen facility in the world capable of
providing information on the electrocouplings of all prominent resonances in the almost unexplored range $Q^2 > 5$~GeV$^2$
\cite{Brodsky:2020vco,burkert2018}.

The three $\gamma_vpN^*$ electrocouplings, $A_{1/2}(Q^2)$, $A_{3/2}(Q^2)$, and $S_{1/2}(Q^2)$, defined in Eqs.~(\ref{Eq:EMWidths1}) 
and (\ref{Eq:EMWidths2}) of Section~\ref{inclusive}, describe the resonance electroexcitation. The longitudinal electrocouplings are 
relevant for electroproduction only, since in photoproduction, owing to gauge invariance of QED, the virtual photon flux of longitudinally 
polarized photons is equal to zero. On the other hand, the behavior of $S_{1/2}$ at $Q^2 < 0.2$~GeV$^2$ is largely unknown and represents 
an interesting open problem to be addressed in electroproduction experiments. The non-resonant contributions to the amplitudes of different
exclusive electroproduction channels are substantially different, while the electrocouplings extracted from different exclusive channels 
should be the same, since the resonance electroexcitation and hadronic decay amplitudes are independent. Therefore, consistent results in 
the extraction of the $\gamma_vpN^*$ electrocouplings from independent studies of different exclusive channels validate reliable extraction 
of these quantities in a nearly model-independent way.

Studies of $N^*$ electroexcitations in different exclusive meson electroproduction channels became feasible with the experimental results from
the CLAS detector, which has provided the dominant part of the available world information on differential cross sections, as well as beam 
$A_b$, target $A_t$, and beam-target $A_{bt}$ asymmetries for most exclusive meson electroproduction channels in the resonance region. These
measurements are summarized in Table~\ref{exclus_el_sum}.

%%%%%%%%%%%%%%%%%%%%%%%%%%%%%%%%%%%%%%%%%%%%%%%%%%%%%%%%%%%%%%%%%%%%%%%%%%%%%%%%%%%%%%%%%%%%%%%%%%%%%%%%%%%%%%%%%%%%%%%%%%%%%%%%%%%%%%%%%%%
\begin{table*}[htb]
\begin{center}
\vspace{2mm}
\begin{tabular}{|c|c|c|c|} \hline
Hadron Final & $W$ Coverage,  & $Q^2$ Coverage, & Measured   \\
State        & GeV            & GeV$^2$        & Observables  \\ \hline
$\pi^+ n$    &  1.1-1.38      & 0.16-0.36  & $\frac{d\sigma}{d\Omega}$  \\
              & 1.1-1.55      & 0.3-0.6    & $\frac{d\sigma}{d\Omega}$    \\
              & 1.1-1.70      & 1.7-4.5    & $\frac{d\sigma}{d\Omega}$, $A_b$    \\ 
              & 1.6-2.00      & 1.8-4.5    & $\frac{d\sigma}{d\Omega}$    \\ \hline
$\pi^0 p$    &  1.1-1.38      & 0.16-0.36  & $\frac{d\sigma}{d\Omega}$  \\
              & 1.1-1.68      & 0.4-1.8    & $\frac{d\sigma}{d\Omega}$, $A_b$, $A_t$, $A_{bt}$    \\
              & 1.1-1.39      & 3.0-6.0    & $\frac{d\sigma}{d\Omega}$   \\ 
              & 1.1-1.80      & 0.4-1.0    & $\frac{d\sigma}{d\Omega}$, $A_b$    \\ \hline 
$\eta p$      & 1.5-2.30      & 0.2-3.1    & $\frac{d\sigma}{d\Omega}$    \\ \hline 
$K^+ \Lambda $      & 1.61-2.60      & 1.40-3.90    & $\frac{d\sigma}{d\Omega}$    \\ 
                    &                & 0.70-5.40    & $P^0$, $P'$    \\ \hline
$K^+ \Sigma^0 $     & 1.68-2.60      & 1.40-3.90    & $\frac{d\sigma}{d\Omega}$    \\ 
                    &                & 0.70-5.40    & $P'$    \\ \hline     
$\pi^+ \pi^- p$     &  1.3-1.6       & 0.20-0.60  & Nine 1-fold  \\
              & 1.4-2.10      & 0.5-1.5    & differential cross    \\
              & 1.4-2.00      & 2.0-5.0    & sections    \\  \hline
\end{tabular}
\end{center}
\caption{Summary of the exclusive meson electroproduction data obtained with the CLAS detector in the resonance region
\cite{fbs-carman,azbu12,mok2019}, including differential cross sections $\frac{d\sigma}{d\Omega}$, beam $A_b$, target $A_t$, and
beam-target $A_{bt}$ asymmetries, as well as recoil $P^0$ and transferred $P'$ polarizations of $\Lambda$ and $\Sigma^0$ hyperons.}
\label{exclus_el_sum}
\end{table*}
%%%%%%%%%%%%%%%%%%%%%%%%%%%%%%%%%%%%%%%%%%%%%%%%%%%%%%%%%%%%%%%%%%%%%%%%%%%%%%%%%%%%%%%%%%%%%%%%%%%%%%%%%%%%%%%%%%%%%%%%%%%%%%%%%%%%%%%%%%%%

A large body of around 150k data points, on differential cross sections, polarization asymmetries, and hyperon recoil/transferred 
polarization has become available spanning almost the entire range of the final state meson emission angles in the CM frame. These data 
provide sufficient information for the extraction of the electrocouplings for most excited nucleon states in the mass range up to 2~GeV for
0.2~GeV$^2 < Q^2 < 5$~GeV$^2$. Tables of the numerical results on the observables listed in Table~\ref{exclus_el_sum} can be found in the 
CLAS Physics Database~\cite{clasdb} and in the SAID database~\cite{saiddb}.

%%%%%%%%%%%%%%%%%%%%%%%%%%%%%%%%%%%%%%%%%%%%%%%%%%%%%%%%%%%%%%%%%%%%%%%%%%%%%%%%%%%%%%%%%%%%%%%%%%%%%%%%%%%%%%%%%%%%%%%%%%%%%%%%%%%%%%%%%%%%%%
\begin{figure}
\centering
  \raisebox{2mm}{\includegraphics[width=0.4\textwidth,height=4.9cm]{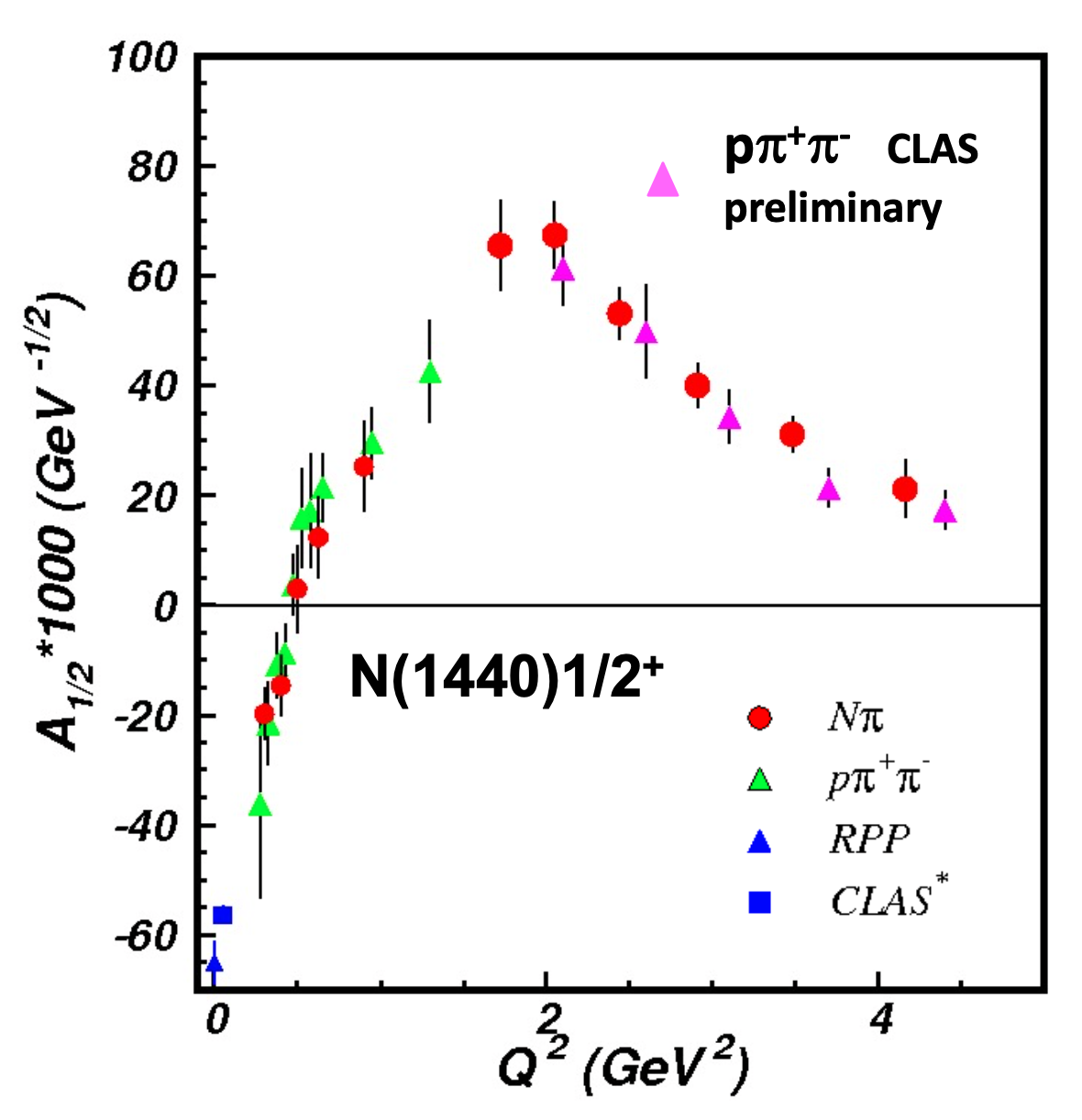}}
  \raisebox{0mm}{\includegraphics[width=0.44\textwidth,height=5.0cm]{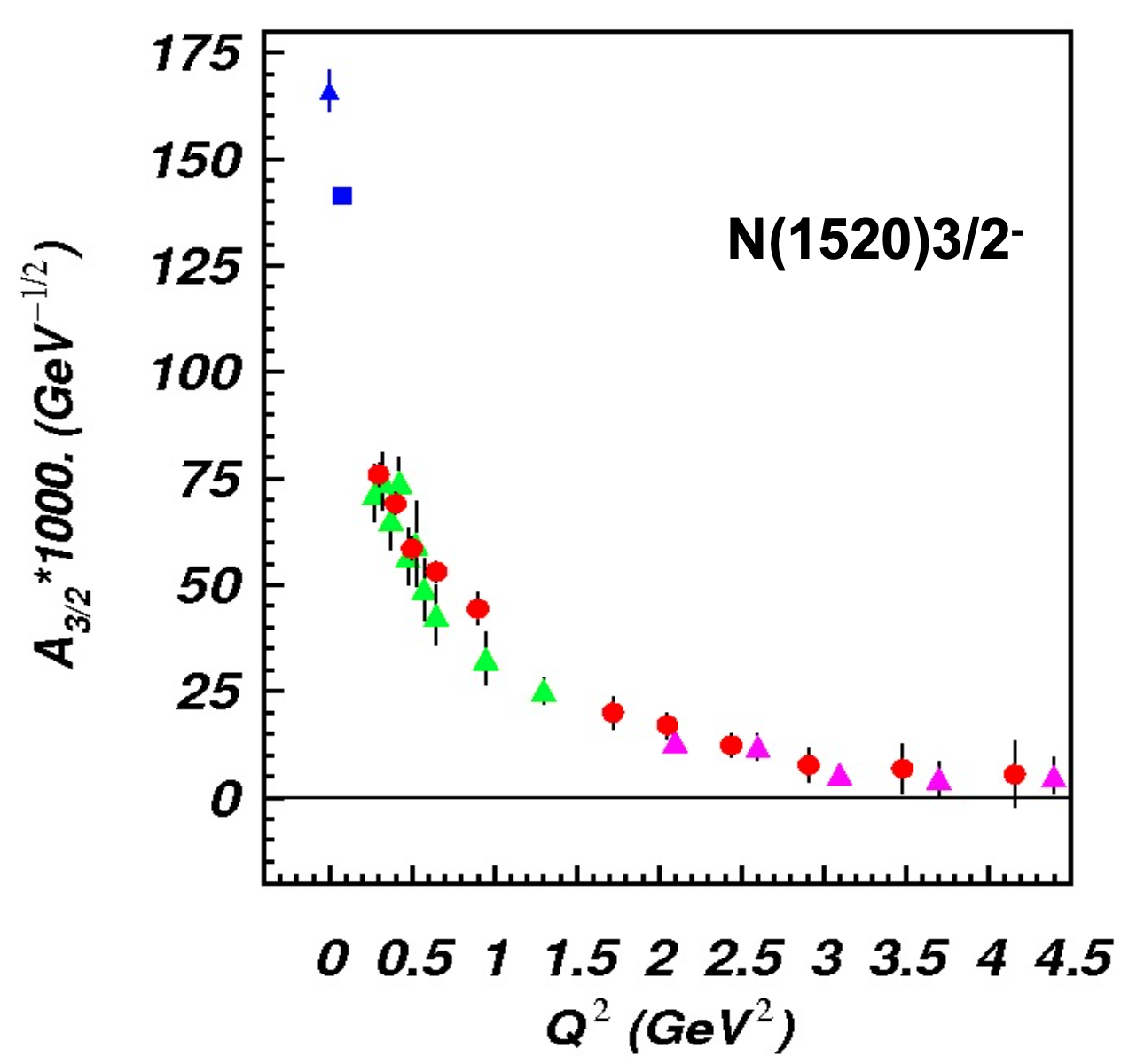}}
\caption{The transverse $\gamma_vpN^*$ photo-/electrocouplings of the $N(1440)1/2^+$ (left) and $N(1520)3/2^-$ (right) determined from 
independent studies of $\pi N$~\cite{aznauryan2009} and $\pi^+\pi^-p$~\cite{mok2020,mokeev2012,mokeev2015} electroproduction off protons.}
\label{1pi2pielcoupl}
\end{figure}
%%%%%%%%%%%%%%%%%%%%%%%%%%%%%%%%%%%%%%%%%%%%%%%%%%%%%%%%%%%%%%%%%%%%%%%%%%%%%%%%%%%%%%%%%%%%%%%%%%%%%%%%%%%%%%%%%%%%%%%%%%%%%%%%%%%%%%%%%%%%

Several approaches were developed for the extraction of the electrocouplings from these data. They include analyses from independent studies 
of individual meson electroproduction channels: $\pi^+ n$ and $\pi^0 p$~\cite{tiator2018,aznauryan2003,aznauryan2009,aznauryan2015,tiator2011},
$\eta p$~\cite{aznauryan2003a,knochlein1995,denizli2007}, and $\pi^+\pi^-p$~\cite{mokeev2009,mokeev2012,mokeev2015}. The results on the
electrocouplings of the $\Delta(1232)3/2^+$ and $N(1440)1/2^+$ resonances have become available from global multi-channel analysis of hadro-,
photo-, and electroproduction data for the first time within the coupled-channel approach developed by the Argonne-Osaka group~\cite{kamano2018}.
Recently, the multipoles for $\pi N$~\cite{mai2021} and $\eta p$~\cite{mai2021a} electroproduction were extracted from the CLAS data at 
$W < 1.6$~GeV and $Q^2 < 5$~GeV$^2$ within the coupled-channel approach developed by the J{\"u}lich-Bonn-Washington group. The 
multipoles inferred from the data pave allow for the extraction of the electrocouplings.

Most of the electrocoupling results obtained so far have become available from independent studies of the CLAS data on $\pi N$ and 
$\pi^+\pi^-p$ electroproduction~\cite{fbs-carman}. The unitary isobar model and dispersion relation approaches developed by the CLAS 
Collaboration to study $\pi N$ electroproduction have provided a good description of the observables in the $W$ range up to 1.7~GeV for 
$Q^2 < 5$~GeV$^2$~\cite{aznauryan2003,aznauryan2009,aznauryan2015}. The data-driven JLab-Moscow (JM) meson-baryon reaction model for
the study of $\pi^+\pi^-p$ electroproduction offers a good description of the data for $W < 2$~GeV and $Q^2 < 5$~GeV$^2$
\cite{mok2019,mokeev2012,mokeev2015}. The reaction models for $\pi N$ and $\pi^+\pi^-p$ electroproduction provide for a reliable isolation of 
the resonant contributions to the observables needed for a reliable extraction of the electrocouplings. A summary of the results from the 
exclusive meson electroproduction data is presented in Table~\ref{electrocoupl_sum}. Currently, the electrocouplings of most proton excited 
states in the mass range $W < 1.8$~GeV and $Q^2 < 5$~GeV$^2$ have become available from the studies of exclusive meson electroproduction with 
CLAS. The most recent results on their values can be found in Ref.~\cite{blin2019}.

%%%%%%%%%%%%%%%%%%%%%%%%%%%%%%%%%%%%%%%%%%%%%%%%%%%%%%%%%%%%%%%%%%%%%%%%%%%%%%%%%%%%%%%%%%%%%%%%%%%%%%%%%%%%%%%%%%%%%%%%%%%%%%%%%%%%%%%%%%%
\begin{table*}[htb]
\begin{center}
\vspace{2mm}
\begin{tabular}{|c|c|c|} \hline
Channel        & Excited Nucleon & $Q^2$ Range (GeV$^2$) of \\
               & States          & Electrocouplings \\ \hline
$\pi^+ n$,     & $\Delta(1232)3/2^+$,       & 0.16-6.0  \\
$\pi^0p$       & $N(1440)1/2^+$, $N(1520)3/2^-$, $N(1535)1/2^-$ & 0.3-4.16 \\ \hline
$\pi^+n$       & $N(1675)5/2^-$, $N(1680)5/2^+$, $N(1710)1/2^+$ & 1.6-4.5 \\ \hline
$\eta p$       & $N(1535)1/2^-$                                 & 0.2-2.6 \\ \hline
$\pi^+ \pi^-p$ & $N(1440)1/2^+$, $N(1520)3/2^-$       & 0.25-1.50, 2.0-5.0 (prelim)   \\
               & $\Delta(1620)1/2^-$, $N(1650)1/2^-$,  &  \\
               & $N(1680)5/2^+$, $\Delta(1700)3/2^-$       & 0.5-1.5  \\
               & $N(1720)3/2^+$, $N'(1720)3/2^+$, $N(1535)1/2^-$ &  \\ \hline
\end{tabular}                        
\end{center}
\caption{The $\gamma_vpN^*$ electrocouplings determined from CLAS data on exclusive meson electroproduction off protons.}
\label{electrocoupl_sum}
\end{table*}
%%%%%%%%%%%%%%%%%%%%%%%%%%%%%%%%%%%%%%%%%%%%%%%%%%%%%%%%%%%%%%%%%%%%%%%%%%%%%%%%%%%%%%%%%%%%%%%%%%%%%%%%%%%%%%%%%%%%%%%%%%%%%%%%%%%%%%%%%%%%

The electrocouplings of the $N(1440)1/2^+$ and $N(1520)3/2^-$ obtained from independent studies of $\pi N$ and $\pi^+\pi^-p$
electroproduction are shown in Fig.~\ref{1pi2pielcoupl}. For both resonances, the electrocouplings determined from these two channels agree
within the error bars. This agreement validates the extraction of these quantities in a nearly model-independent way. This success has
also validated the reaction models~\cite{aznauryan2003,aznauryan2009,aznauryan2015,mokeev2009,mokeev2012,mokeev2015} used to extract these
quantities. The CLAS measurements~\cite{isupov2017,trivedi2018} have extended the kinematic coverage of the $\pi^+\pi^-p$ electroproduction 
data for 1.4~GeV $< W < 2.0$~GeV and 2.0~GeV$^2 < Q^2 < 5.0$~GeV$^2$. Extraction of the electrocouplings of most excited states of the proton 
in this kinematic range from these data is in progress. The expected results will provide valuable information on the evolution of the 
resonance structure in the transition from the interplay between the inner quark core and the external meson-baryon cloud to the regime where 
the quark core dominates~\cite{fbs-carman}. Recently, new data from CLAS on the $\pi^0p$ differential cross sections~\cite{markov2020} and 
beam asymmetries~\cite{isupov2021} have become available. The sensitivity of these observables to the contributions of the nucleon resonances 
in the third resonance region was demonstrated. The upcoming results on the electrocouplings of the nucleon resonances in the third resonance 
region from these $\pi N$ data, as well as from the $\pi^+\pi^-p$ electroproduction data available within the same kinematic range
\cite{fedotov2018}, will shed further light on the interplay between the quark core and the meson-baryon cloud in the structure of excited 
nucleon states~\cite{Burkert:2019bhp,azbu12,mokeev2015}. 

In the mass range $W > 1.6$~GeV, several nucleon resonances decay preferentially to $\pi\pi N$ with a small branching fraction to $\pi N$. 
This makes $\pi^+\pi^-p$ electroproduction the major source of information on the electrocouplings of these states
\cite{mok2020,mokeev2020a,mok2019}. These quantities can also be determined independently from exclusive $K\Lambda$ and $K\Sigma$ ($KY$)
electroproduction. The quality of the CLAS data on $KY$ electroproduction~\cite{5st,carman13,carman03,rauecarman,sltp,carman09,ipol} certainly
provides an opportunity to extract the electrocouplings from these exclusive channels when a suitable reaction model becomes available. The
development of such a reaction model for the extraction of the electrocouplings from the available $KY$ electroproduction data represents an
urgent need for advancing the exploration of the spectrum and structure of the higher-lying nucleon resonances. The exclusive $\pi^+\pi^-p$ 
and $KY$ channels look very promising in the search for new baryon states in the combined studies of meson photo- and 
electroproduction data~\cite{bur2020,fbs-carman}.

\section{Nucleon Resonance Electrocouplings and Emergence of Hadron Mass}

The emergence of hadron mass represents a challenging open problem in the Standard Model that arises from the comparison between the 
measured masses of protons and neutrons and the sum of the masses of their current-quark constituents. Nucleons consist of the lightest $u$ 
and $d$ quarks. By adding the masses of the current quarks in the proton and neutron, which enter into the QCD Lagrangian and are generated 
through the Higgs mechanism, we get the current quark mass contribution to the proton of $8.99^{+1.45}_{-0.65}$~MeV and to the neutron of
$11.50^{+1.45}_{-1.60}$~MeV~\cite{pdg}. The sum of the current quark masses accounts for $<$1.1\% and $<$1.4\% of the measured proton and 
neutron masses, respectively. Important questions arise, namely, which mechanism underlies the generation of the dominant part of $>$98\% of 
hadron mass and how it is related to the generation of the non-zero trace of the nucleon energy-momentum tensor, the so-called trace anomaly? 

The impressive progress achieved using CSMs during the past decade has conclusively demonstrated that the dominant part of hadron mass is 
generated by the strong interaction in the regime of large QCD running coupling through the dressing of the bare QCD quarks and gluons driven 
by the QCD Lagrangian~\cite{roberts2021,roberts2020b}. Currently, CSMs present the only available approach capable of making predictions on the 
EHM manifestation in the $Q^2$ evolution of the pion and nucleon elastic form factors, the pion parton distribution function (PDF), and the
$\gamma_vpN^*$ electrocouplings within a common framework for both mesons and baryons. The studies of electrocouplings play an important 
role in validating the EHM concept developed within CSMs from the experimental data~\cite{Brodsky:2020vco,Burkert:2019bhp,fbs-carman}.

%%%%%%%%%%%%%%%%%%%%%%%%%%%%%%%%%%%%%%%%%%%%%%%%%%%%%%%%%%%%%%%%%%%%%%%%%%%%%%%%%%%%%%%%%%%%%%%%%%%%%%%%%%%%%%%%%%%%%%%%%%%%%%%%%%%%%%%%%%%%%
\begin{figure*}
\begin{center}
  \raisebox{0mm}{\includegraphics[width=1.0\textwidth,height=6.0cm]{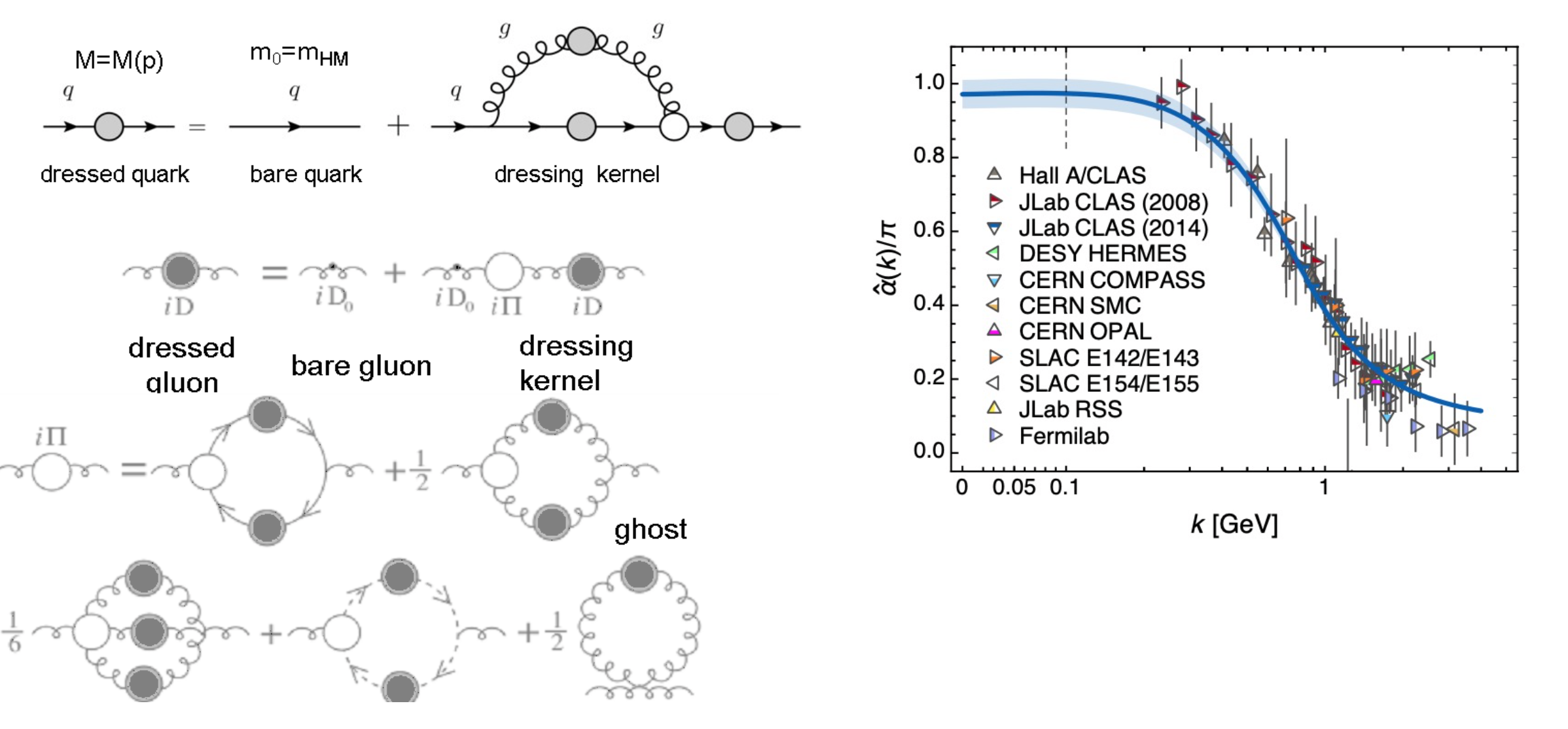}}
\vspace{-4mm}
\caption{(Left) Graphical representation of the QCD equations of motion for the quark and gluon fields that underlie the generation of 
the dressed quark and gluon masses through dynamical chiral symmetry breaking. (Right) QCD running coupling evaluated using CSMs 
starting from the QCD Lagrangian~\cite{cui20,binosi2017,roberts2020c}.}
\label{qg_dressing}
\end{center}
\end{figure*}
%%%%%%%%%%%%%%%%%%%%%%%%%%%%%%%%%%%%%%%%%%%%%%%%%%%%%%%%%%%%%%%%%%%%%%%%%%%%%%%%%%%%%%%%%%%%%%%%%%%%%%%%%%%%%%%%%%%%%%%%%%%%%%%%%%%%%%%%%%%%%

%%%%%%%%%%%%%%%%%%%%%%%%%%%%%%%%%%%%%%%%%%%%%%%%%%%%%%%%%%%%%%%%%%%%%%%%%%%%%%%%%%%%%%%%%%%%%%%%%%%%%%%%%%%%%%%%%%%%%%%%%%%%%%%%%%%%%%%%%%%%%
\begin{figure}
\centering
  \raisebox{0mm}{\includegraphics[width=0.45\textwidth,height=5.2cm]{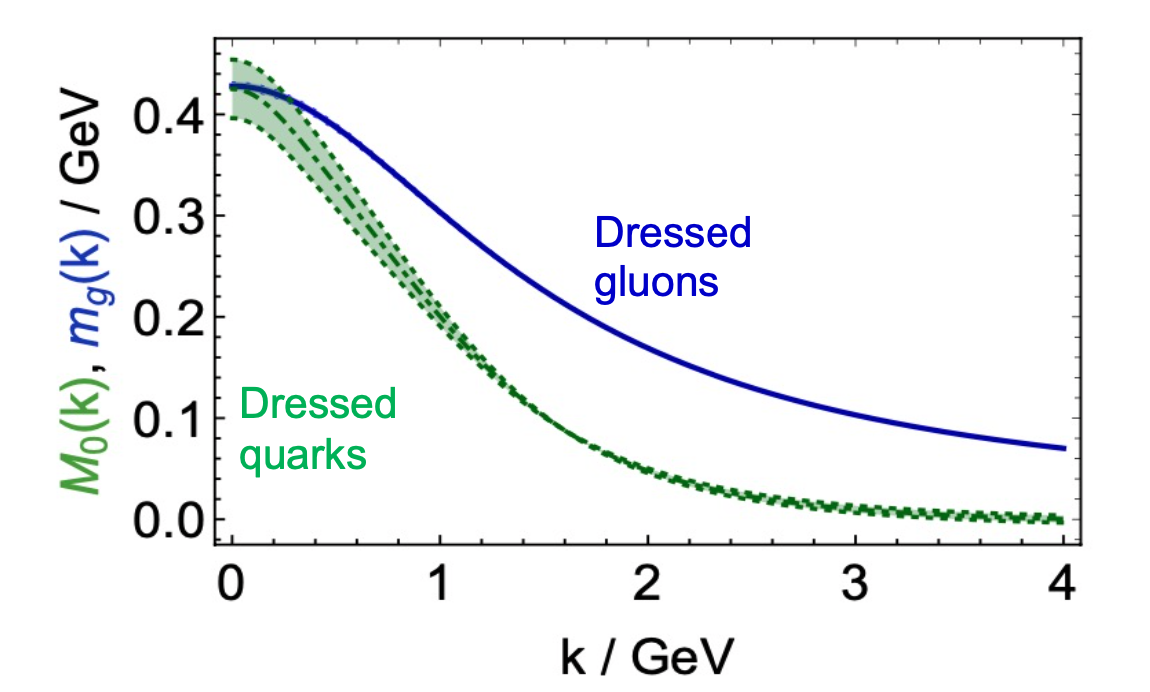}}
  \raisebox{0mm}{\includegraphics[width=0.5\textwidth,height=5.2cm]{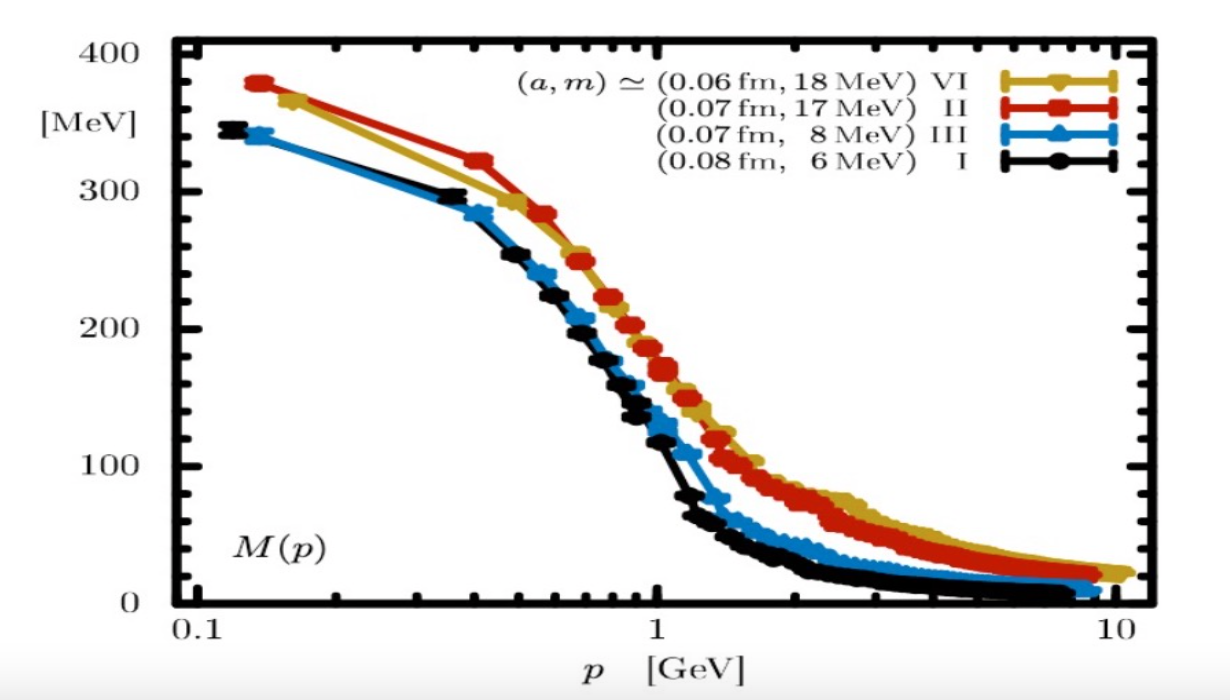}}
\vspace{-2mm}
\caption{(Left) Momentum dependence of the $u$ and $d$ dressed quark masses computed with CSMs starting from the QCD Lagrangian
\cite{roberts2020c}. (Right) Momentum dependence of the $u$ and $d$ dressed quark masses evaluated within LQCD for different lattice 
spacings $a$ and current quark masses $m$~\cite{oliviera2019}.}
\label{run_masses}
\end{figure}
%%%%%%%%%%%%%%%%%%%%%%%%%%%%%%%%%%%%%%%%%%%%%%%%%%%%%%%%%%%%%%%%%%%%%%%%%%%%%%%%%%%%%%%%%%%%%%%%%%%%%%%%%%%%%%%%%%%%%%%%%%%%%%%%%%%%%%%%%%%%

The QCD Lagrangian implies that the current quarks should be dressed in the processes shown in the top left part of Fig.~\ref{qg_dressing}.
Similarly, the gauge QCD gluons should be dressed in the processes shown in the bottom left part of Fig.~\ref{qg_dressing}
\cite{cui20,binosi2017,roberts2020c}. In the regime of large QCD running coupling (see Fig.~\ref{qg_dressing} (right)), the QCD-driven 
dressing processes shown in Fig.~\ref{qg_dressing} (left) generate dressed quarks and gluons with dynamical distance/momentum-dependent 
masses. The dressed quark and gluon masses evaluated using CSMs from the QCD Lagrangian are shown in Fig.~\ref{run_masses} (left)
\cite{roberts2020c}. The dressed quark mass function shows how almost massless, perturbative QCD quarks in the ultraviolet at momenta 
$k > 2$~GeV, eventually in the infrared at momenta $k < 0.5$~GeV, become fully dressed quarks of $\approx$400~MeV mass that are employed in 
the constituent quark models. In the infrared regime, the dressed quark and gluon running masses converge, setting up the strong interaction 
mass scale of $\approx$400~MeV, which is consistent with the measured nucleon mass. The Higgs boson discovered at CERN accounts for the 
generation of only the current quark mass, with a contribution into the fully dressed quark mass of $<$2\%. The Higgs mechanism is almost 
irrelevant for the generation of mass of the nucleon and its excited states. The dominant part of the $N/N^*$ masses is generated through 
the quark and gluon dressing processes shown in Fig.~\ref{qg_dressing} (left) that are responsible for dynamical chiral symmetry breaking 
(DCSB). In the strong QCD regime, the energy stored in the gluon field is transferred into the running mass of the dressed quarks. The 
interaction between the three dressed quarks with dynamically generated masses eventually produces the experimentally observed masses of 
the ground and excited states of the nucleon. Notably, the CSM results on the momentum dependence of the dressed quark mass are fully 
consistent with the independent evaluation of this quantity within lattice QCD (LQCD) simulations~\cite{oliviera2019}. The LQCD results are 
shown in Fig.~\ref{run_masses} (right) and confirm the CSM predictions. The dressed quark mass function is computed with the CSM as the 
solution of the QCD equations of motion for the quark and gluon fields in the strong QCD regime. So far, these equations can be solved under
approximations mostly related to the description of the dressed quark-gluon vertex. It is a challenging task for contemporary experimental 
hadron physics to check the CSM concept on EHM. A large body of predictions on both meson and baryon structure observables was made within 
the CSM, offering an excellent opportunity to check the CSM concept for EHM against different sets of hadron structure observables
\cite{Brodsky:2020vco,fbs-carman,roberts2020c}.

Studies of the $\gamma_vpN^*$ electrocouplings, in combination with the nucleon elastic form factors, play a particular role in this effort,
allowing for the exploration of EHM in conditions when the dominant part of the ground and excited nucleon state mass comes dynamically 
from the generated dressed quark masses. Consistent results on the dressed quark mass function from independent studies of nucleon elastic 
form factors and the electrocouplings of nucleon resonances of different structure provides insight into EHM in a nearly model-independent way. 

%%%%%%%%%%%%%%%%%%%%%%%%%%%%%%%%%%%%%%%%%%%%%%%%%%%%%%%%%%%%%%%%%%%%%%%%%%%%%%%%%%%%%%%%%%%%%%%%%%%%%%%%%%%%%%%%%%%%%%%%%%%%%%%%%%%%%%%%%%%%
\begin{figure*}
\begin{center}
  \raisebox{0mm}{\includegraphics[width=0.88\textwidth,height=5.0cm]{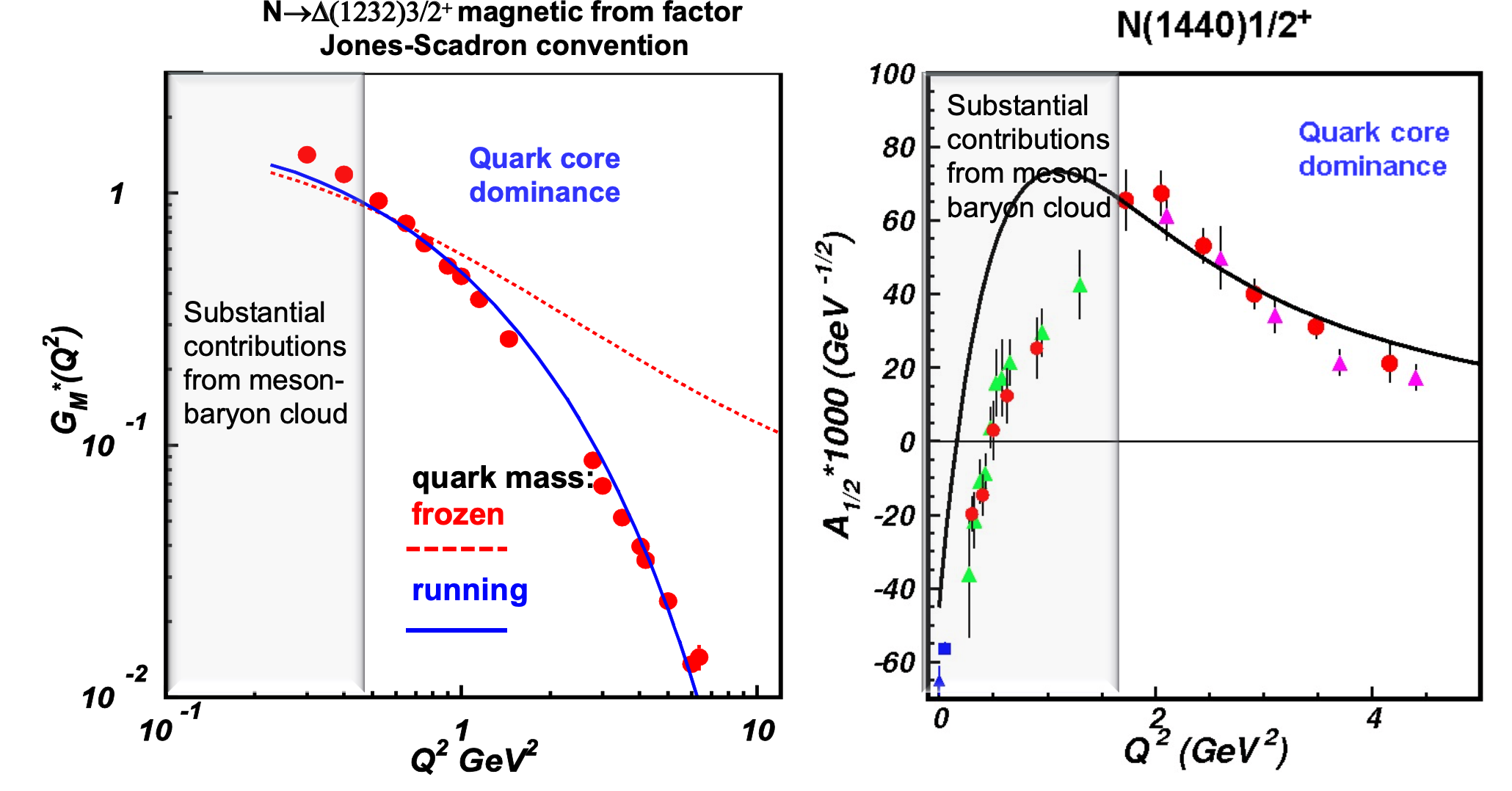}}
\vspace{0mm}
\caption{(Left) $N \to \Delta(1232)3/2^+$ magnetic transition form factor in the Jones-Scadron convention in comparison with CSM predictions
\cite{segovia2014}. The CSM prediction for the contact $qq$-interaction with a frozen quark mass is shown by the red dashed line, while the 
CSM prediction with a realistic $qq$-interaction that results in a running quark mass is shown by the blue solid line. (Right) $A_{1/2}$
electrocoupling of the $N(1440)1/2^+$ resonance from the CLAS measurements in comparison with the CSM prediction~\cite{segovia2015} shown by 
the black solid line.}
\label{roper_delta}
\end{center}
\end{figure*}
%%%%%%%%%%%%%%%%%%%%%%%%%%%%%%%%%%%%%%%%%%%%%%%%%%%%%%%%%%%%%%%%%%%%%%%%%%%%%%%%%%%%%%%%%%%%%%%%%%%%%%%%%%%%%%%%%%%%%%%%%%%%%%%%%%%%%%%%%%%%

The CLAS results on the electrocouplings of low-lying nucleon resonances in the mass range $W < 1.65$~GeV have already demonstrated a 
profound impact on the understanding of EHM. CSMs have, for the first time, provided a good description of the CLAS results on 
the electrocouplings of the $\Delta(1232)3/2^+$ and $N(1440)1/2^+$ starting from the QCD Lagrangian (see Fig.~\ref{roper_delta})
\cite{chen2019,segovia2016,segovia2016a,segovia2016b,segovia2015,segovia2014}. The CSM predictions for the dominant $N \to \Delta(1232)3/2^+$
magnetic transition form factor were obtained under two assumptions for the $qq$-interaction. The exploratory computation of this form
factor employed a simplified contact $qq$-interaction~\cite{segovia2014}. Such an interaction allows for the description of the dressed quark
mass of 360~MeV, but results in a momentum-independent ``frozen" dressed quark mass function (see the dashed red line in 
Fig.~\ref{roper_delta} (left)). This CSM prediction overestimates the experimental results at $Q^2 > 1$~GeV$^2$. The discrepancy between the 
data and the CSM prediction increases with $Q^2$, approaching an order magnitude at $Q^2 \approx 6$~GeV$^2$. Instead, use of a realistic
$qq$-interaction~\cite{roberts2020c}, which predicts a momentum-dependent dressed quark mass (see Fig.~\ref{run_masses}), offers a good 
description of the CLAS results in the entire range of $Q^2 > 0.8$~GeV$^2$, where the contribution from the quark core into the structure of
the $\Delta(1232)3/2^+$ becomes the largest. Analysis of the CLAS results on the $N \to \Delta(1232)3/2^+$ magnetic transition form factor 
using CSMs has conclusively demonstrated for the first time that the dressed quark mass is, in fact, running. 

Significantly, a good description of both the $\Delta(1232)3/2^+$ and $N(1440)1/2^+$ electrocouplings (see Fig.~\ref{roper_delta}) was achieved 
with exactly the {\it same} dressed quark mass function as was used previously for the successful description of the pion and nucleon elastic 
form factors and the pion PDF~\cite{roberts2021,roberts2020c,segovia2014}. Consistent results on the momentum dependence of the dressed quark 
mass obtained from independent experimental studies of the structure of the ground state pion and nucleon, as well as from the results on 
the $Q^2$ evolution of the resonance electrocouplings of different excited nucleon states of different structure, the first spin-isospin flip 
for the $\Delta(1232)3/2^+$, and the first radial excitation of the $N(1440)1/2^+$, strongly suggest that dressed quarks with a dynamically
generated mass represent an active component in the structure of the ground state pion and nucleon, as well as the $\Delta(1232)3/2^+$ and
$N(1440)1/2^+$ resonances. This success offers strong support for the CSM concept of EHM from the experimental data and demonstrates the 
capability to gain insight into EHM from the experimental data on the pion and the ground/excited states of the nucleon, analyzed within the 
common CSM theoretical framework. This is a very impressive achievement in hadron physics in the past decade resulting from synergistic efforts
between experiment, phenomenology, and QCD-rooted hadron structure theory.

%%%%%%%%%%%%%%%%%%%%%%%%%%%%%%%%%%%%%%%%%%%%%%%%%%%%%%%%%%%%%%%%%%%%%%%%%%%%%%%%%%%%%%%%%%%%%%%%%%%%%%%%%%%%%%%%%%%%%%%%%%%%%%%%%%%%%%%%%%%%
\begin{figure*}
\begin{center}
  \raisebox{0mm}{\includegraphics[width=0.48\textwidth,height=5.0cm]{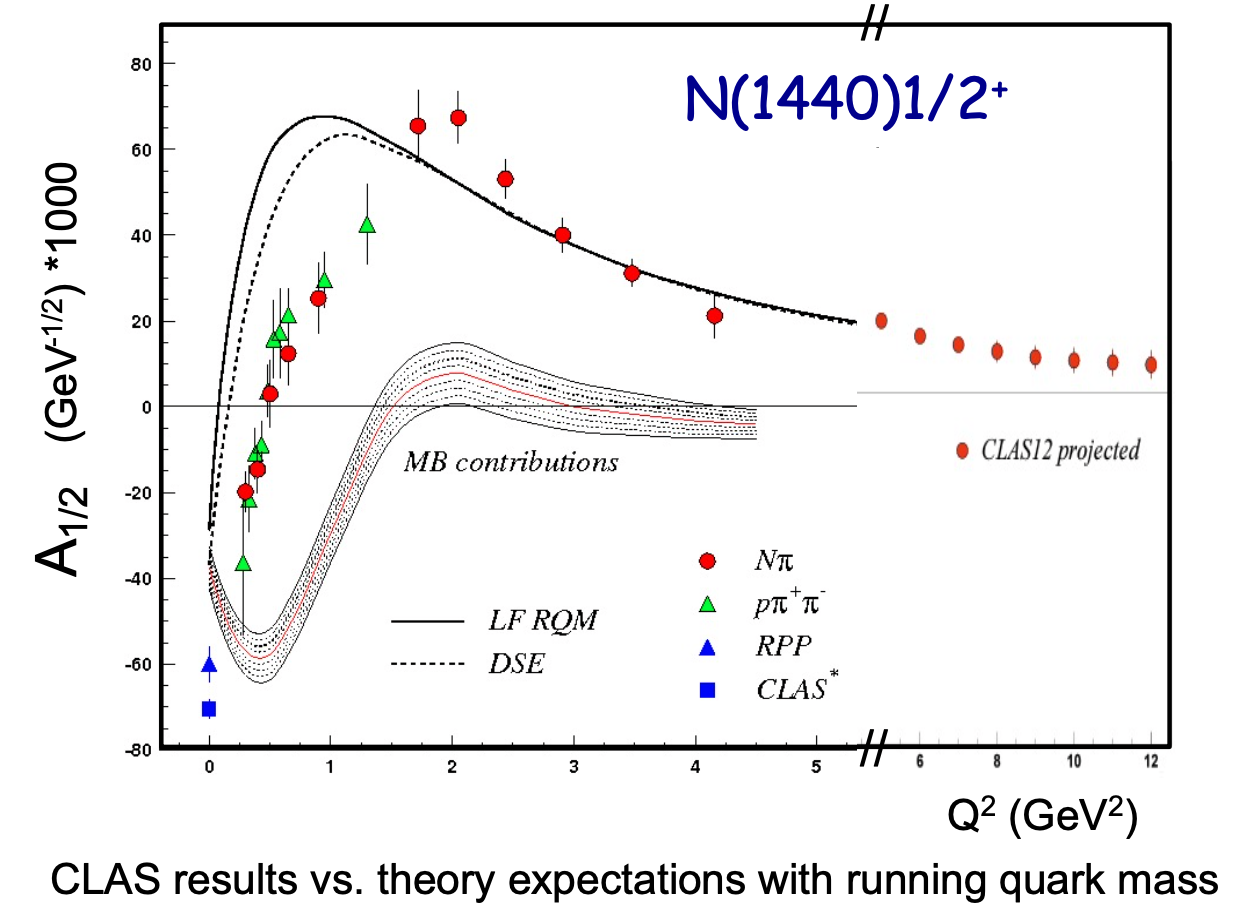}}
\raisebox{5.8mm}{\includegraphics[width=0.48\textwidth,height=5.0cm]{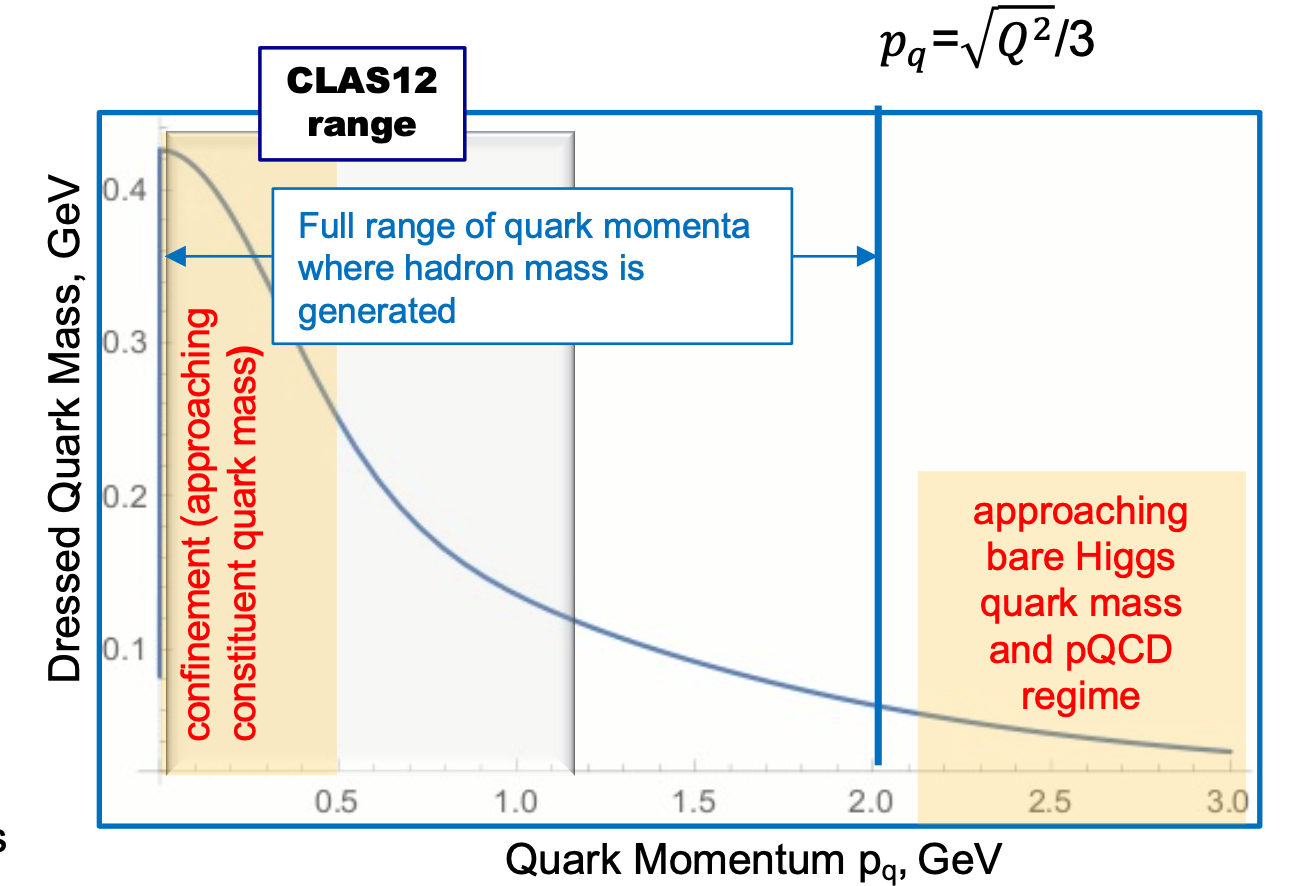}}  
%\vspace{-2mm}
\caption{(Left) Expected results on the $N(1440)1/2^+$ electrocouplings from CLAS12 data in comparison with those available from
CLAS~\cite{fbs-carman}. (Right) Kinematic coverage for insight into the dressed quark mass function from the electrocoupling data with
CLAS~\cite{fbs-carman} (left yellow shaded area), CLAS12~\cite{Brodsky:2020vco}, and from the foreseen JLab measurements with increased 
energy up to 24~GeV and luminosity above $10^{36}$~cm$^{-2}$s$^{-1}$ as a part of the future JLab research program (area between the vertical 
lines).}
\label{clas12_nstar}
\end{center}
\end{figure*}
%%%%%%%%%%%%%%%%%%%%%%%%%%%%%%%%%%%%%%%%%%%%%%%%%%%%%%%%%%%%%%%%%%%%%%%%%%%%%%%%%%%%%%%%%%%%%%%%%%%%%%%%%%%%%%%%%%%%%%%%%%%%%%%%%%%%%%%%%%%%

The predictions of the electrocouplings of the $\Delta(1600)3/2^+$ have recently become available from CSMs~\cite{lu2019}. The extraction
of the electrocouplings for this state from the CLAS $\pi^+\pi^-p$ electroproduction data~\cite{isupov2017,trivedi2018} at 
2.0~GeV$^2 < Q^2 < 5.0$~GeV$^2$ is in progress. A successful description of these quantities will solidify insight into the momentum 
dependence of the dressed quark mass in a nearly model-independent way. CSMs continue to extend the exploration of $N^*$ 
structure with the first results on the electrocouplings of the $N(1535)1/2^-$ resonance~\cite{raya2021}. In leading order, the structure of 
this state can be described as an $L=1$ orbital excitation of three dressed quarks. Systematic studies of the resonances that belong to the
[$70,1^-$] SU(6) spin-flavor supermultiplet will shed light on either the universality or the environmental sensitivity of the 
dressed quark mass function.

Currently, the CLAS12 detector in Hall~B at JLab is the only available and foreseen facility in the world capable of exploring exclusive 
meson electroproduction in the resonance region at the highest $Q^2 > 5$~GeV$^2$ ever achieved in the studies of exclusive electroproduction. 
The electrocouplings of all prominent nucleon resonances will be determined from the data of experiments with CLAS12 within the $Q^2$ range 
from 5~GeV$^2$ to 10~GeV$^2$ from independent and combined studies of the $\pi N$, $K\Lambda$, $K\Sigma$, and $\pi^+\pi^-p$ channels
\cite{Brodsky:2020vco,fbs-carman}. As a representative example, the quality of the expected results on the electrocouplings of the 
$N(1440)1/2^+$ in comparison with the available results is shown in Fig.~\ref{clas12_nstar} (left). The $Q^2$ coverage of CLAS12 will allow 
for the mapping out of the dressed quark mass function in the range of quark momenta up to 1.3~GeV. This is an essential extension in 
comparison to the range of quark momentum $<$0.5~GeV accessible from the CLAS results (see Fig.~\ref{clas12_nstar} (right)). Consistent 
results on the momentum dependence of the dressed quark mass function from the studies of the electrocouplings over the full spectrum of the 
excited nucleon states of distinctively different structure, different spin-isospin flips, and different radial and orbital excitations will
validate credible insight into this fundamental quantity in a nearly model-independent way. The expected CLAS12 results on the electrocouplings 
of most excited nucleon states will allow the key open problems in the Standard Model on the emergence of hadron mass and quark gluon 
confinement to be addressed.

In order to overlap the full range of dressed quark momenta where the dominant part of hadron mass is generated (see 
Fig.~\ref{clas12_nstar} (right)), the electrocouplings should be determined within the range of $Q^2$ up to 36~GeV$^2$, assuming roughly 
equal sharing of the momentum transferred by virtual photons among the three dressed quarks. Such experiments require a further increase of 
the JLab electron beam energy up to 24~GeV and the construction of a new large-acceptance detector capable of measuring exclusive meson 
electroproduction at luminosities above 10$^{36}$~cm$^{-2}$s$^{-1}$. The exploration of EHM at $Q^2 > 10$~GeV$^2$ motivates the future 
experimental studies of resonance structure after the completion of the 12-GeV program.

\section{Conclusion and Outlook}

Studies of exclusive meson photoproduction in the resonance region have provided precise results on the differential cross sections and 
polarization observables for most exclusive photoproduction channels. Progress in the amplitude analyses has allowed for the establishment
of the masses, total/partial hadronic decay widths, and photocouplings for all prominent resonances in the mass range $<$2.5~GeV. Further 
efforts are needed to improve agreement between the $N^*$ parameters available from different analysis approaches. Several long-awaited new 
baryon states (referred to as ``missing" resonances) have been discovered in global multi-channel analyses of exclusive meson photo- and
hadroproduction data with a major impact from the CLAS $K\Lambda$ and $K\Sigma$ photoproduction data. The new $N'(1720)3/2^+$ resonance has 
been observed in the combined studies of $\pi^+\pi^-p$ photo- and electroproduction data measured with CLAS. Currently, this is the only new 
resonance for which the results on the $Q^2$-evolution of the electrocouplings have become available, allowing us to gain insight into the 
structure of new baryon states. The experiments of the 12-GeV era at JLab, as well as at facilities in Asia and Europe, will complete studies 
of the spectrum of the excited nucleon states, including the search for hybrid baryons with glue as an active structural component. These 
studies will address approximate symmetries for the strong QCD regime that underlie the generation of the $N^*$ spectrum, the emergence of 
hadrons from a deconfined mixture of quarks and gluons in the early Universe, and the dual role of gluons as the strong force carrier and 
as hadron matter constituents. Experiments of the 12-GeV era at JLab will also considerably extend our knowledge on the spectra of excited 
baryons with strange quarks~\cite{burkert2018,klong2020}.

The CLAS detector has provided the dominant part of the world data on most exclusive meson electroproduction channels in the resonance region.
These data have allowed for the determination of the electrocouplings of most resonances in the mass range up to 1.8~GeV with consistent 
results from analyses of the $\pi^+n$, $\pi^0p$, $\eta p$, and $\pi^+\pi^-p$ channels. The resonance electrocouplings will become available for 
most nucleon resonances in the mass range $<$2~GeV and at $Q^2 < 5$~GeV$^2$ from CLAS and at $Q^2 < 10$~GeV$^2$ from CLAS12. The EHM paradigm 
makes a broad array of predictions for the structure of $N/N^*$ states that are worth testing to gain insight and understanding of hadron mass
generation by mapping out the momentum dependence of the dressed quark running masses. A good description of the CLAS results on the
$\Delta(1232)3/2^+$ and $N(1440)1/2^+$ electroexcitation amplitudes achieved with the same dressed quark mass function as used previously in
the successful evaluations of the nucleon elastic and pion form factors and the pion PDF, validate insight into the dynamics that underlie the 
emergence of hadron mass. Studies of the $\Delta(1600)3/2^+$ electrocouplings are in progress and, when complete, may be compared with CSM
predictions made using the same dressed quark mass function employed in all other calculations. CLAS12 is the only facility in the world 
capable of obtaining the electrocouplings of all prominent $N^*$ states at the still unexplored ranges at the highest $Q^2$ for exclusive 
reactions from 5~GeV$^2$ to 10~GeV$^2$ from measurements of $\pi N$, $KY$, and $\pi^+\pi^-p$, allowing us to map out the dressed quark mass 
function at quark momenta up to 1.3~GeV. Extension of the results on the electrocouplings into the $Q^2$ range from 10~GeV$^2$ to 36~GeV$^2$ 
from measurements at future facilities with luminosity $>$10$^{36}$~cm$^{-2}$s$^{-1}$ will allow for the completion of the exploration of 
the dressed quark mass function within the full range of distances where the bulk of hadron mass is generated in the transition from 
quark-gluon confinement to the regime of perturbative QCD. Studies over this full range are essential in order to address the most 
challenging problems of the Standard Model on the nature of $>$98\% of the mass of hadrons.

\section*{Acknowledgments}

This work was supported in part by the U.S. Department of Energy and the Skobeltsyn Nuclear Physics Institute and Physics
Department at Lomonosov Moscow State University. The Southeastern Universities Research Association (SURA) operates the Thomas Jefferson 
National Accelerator Facility for the U.S. Department of Energy under Contract No. DE-AC05-06OR23177.



\end{document}